\documentclass[a4,10pt,onecolumn,oneside,notitlepage,final]{article}

\usepackage{bm}

\def \in #1 #2 {\int \limits_{#1}^{#2}}
\def\Lag{\mathcal{L}} 

\usepackage{amssymb}
\usepackage{amsmath}
\usepackage{graphicx} 
\usepackage{color}
\usepackage{axodraw}

\usepackage{amsmath}

\usepackage{pstricks}

\def\localinput#1{{
  \renewcommand{\documentclass}[2][dummy]{}
  \renewcommand{\usepackage}[2][dummy]{}
  \renewenvironment{document}{}{}
  \def\jobname{#1}
  \input{#1}
}}

\def\sla#1{\ooalign{\hfil\hspace{-0.1ex}\raise.2ex\hbox{$\not \phantom{#1}$}\hfil\crcr  $#1$}}

\def\siml{\hspace{0.3em}\raisebox{0.4ex}{$<$}\hspace{-0.75em}\raisebox{-.7ex}{$\sim$}\hspace{0.3em}}

\begin{document}

\begin{titlepage}

\date{}

\begin{flushright}
RUP-08-2
\end{flushright}
\vspace{12ex}
\begin{center}
 \huge{Lepton Flavor Violating $\tau\to3\mu$ Decay \\in \\Type-I\!I\!I Two Higgs Doublet Model}

\vspace{3ex}

\large{\hspace{-0em}  Akihiro Matsuzaki\footnote{akihiro@rikkyo.ac.jp}
 \ and Hidekazu Tanaka\footnote{tanakah@rikkyo.ac.jp}
\\[4mm]
 \small\textit{
 Department of Physics, Rikkyo University,}
 \\ \small\textit{Nishi-ikebukuro, Toshima-ku Tokyo, Japan, 171  }}

\vspace{3ex}

%\hfil{\Huge  File Name:\textcolor{red}{ \jobname.tex}} \hfil

\end{center}

%\section{abstract}
\begin{abstract}
    We study the current structure of the lepton flavor violating $\tau\to3\mu$ decay in Type-I\!I\!I 2HDM.
    This model has many coupling constants which affect this decay.
    We find that each coupling constant corresponds to the different final-state momenta distribution and vice versa. 
    Using this fact, we suggest how to determine the current structure. 
    We also find the upper limit $|\eta_{23}^E\eta_{22}^E|<0.00022$ in the case that all Higgs bosons except for the lighter CP even neutral one $h^0$ are decoupled, $M_{h^0}=115$GeV and $\cos\beta=1/\sqrt{2}$.     
    The observable difference between the MSSM and type-I\!I\!I 2HDM is also discussed.
\end{abstract}

\end{titlepage}

\section{Introduction}
    In the Standard Model (SM), which is supported by many experimental data, only the Higgs boson is undiscovered.
    The Large Hadron Collider (LHC) will start and search it \cite{LHC}.
    In LHC, we hope to discover many new particles since they are very important hints to beyond the SM.
    However these particles may be too heavy to discover in LHC.
    Even if so, the Higgs boson mass has the upper limit from the unitarity \cite{Unitarity}.  
    In the Higgs sector, many models are suggested e.g. minimal supersymmetric SM (MSSM), little Higgs, technicolor and two Higgs doublet model (2HDM) \cite{MSSM}, \cite{LH}, \cite{TC}, \cite{Hhg}.
    Especially, 2HDM can be an effective theory of the models defined at higher energy scale.
    The purposes of this paper are  
\begin{enumerate}
  \item the determination of the theory in Higgs sector when Higgs boson(s) is (are) discovered in LHC, and
  \item the determination of the current structure in Type-I\!I\!I 2HDM. 
\end{enumerate}
    If only one neutral Higgs boson is discovered, is that means that the Higgs sector is the SM?
    The same situation occurs in MSSM and 2HDM if other four Higgs bosons are decoupled since very heavy or weakly coupled. 
    Even if we can determine that the Higgs sector is 2HDM-like by LHC experiment, this model has many currents and couplings, and each coupling constant is a complex parameter of the model.
    It is very important to determine the absolute values of coupling constants, and relative phases between them.

    The lepton flavor violation (LFV) process gives them an answer.
    These models beyond the SM predict the large LFV \cite{Saku}, \cite{Babu}, \cite{2HDM1}.
    The coupling constant between Higgs boson and fermions tends to be larger proportional to the fermion mass.  
    Especially, the Type-I\!I\!I 2HDM has tree-level flavor changing neutral currents (FCNCs).
    Moreover, KEK B-factory generates huge number of $\tau^+ \tau^-$ pairs.  
    These facts suggest LFV $\tau$ decay may appear in near future.  
    In this paper, we study $\tau\to3\mu$ mode.
    The reasons are as follows:
\begin{itemize}
  \item the Higgs boson can contribute in tree level,
  \item we expect the clear experimental result since the final state has no photon and no missing particle and 
  \item we can study the polarization information using the initial and final energy momentum distributions \cite{Okada},\cite{Matsuzaki}. 
\end{itemize}

    The general Lagrangian for $\tau^+\to \mu^+ \mu^+ \mu^-$ decay is written as  \cite{Okada}, \cite{Matsuzaki}:
\begin{align} \begin{split} \label{Lag}
\Lag&=- 2\sqrt{2}G_F \Bigr\{
                \hspace{1em}        g_{1}(\bar{\tau}_R \mu_L)(\bar{\mu}_R  \mu_L)
            +g_{2}(\bar{\tau}_L \mu_{R})(\bar{\mu}_L  \mu_R)  \\
 &   \hspace{8em}     +g_{3}(\bar{\tau}_R \gamma_\alpha \mu_R)(\bar{\mu}_R \gamma^\alpha  \mu_R)
            +g_{4}(\bar{\tau}_L \gamma_\alpha \mu_L)(\bar{\mu}_L \gamma^\alpha  \mu_L)   \\
 &   \hspace{8em}   +g_{5}(\bar{\tau}_R \gamma_\alpha \mu_R)(\bar{\mu}_L \gamma^\alpha  \mu_L)
            +g_{6}(\bar{\tau}_L \gamma_\alpha \mu_L)(\bar{\mu}_R \gamma^\alpha  \mu_R) \Bigl\},    
  \\   &  - 2\sqrt{2}G_F m_\tau\Bigr\{
              A_R\bar{\tau}_R\sigma^{\alpha\beta}\mu_L F_{\alpha\beta}
            + A_L\bar{\tau}_L\sigma^{\alpha\beta}\mu_R F_{\alpha\beta}\Bigl\}  \\
         &\hspace{5em} +\bar{\mu} (i D^\alpha\gamma_\alpha -m_\mu)  \mu         
              -\frac{1}{4}   F^{\alpha\beta}F_{\alpha\beta},
    \end{split} \end{align}
    where  $m_\mu$ and $m_\tau$ are the masses of the $\mu^\pm$ and $\tau^\pm$, respectively; $G_F $ is the Fermi constant; $ \{ \bar{\tau}_L, \bar{\mu}_L,  \mu_R\}$ and $\{\bar{\tau}_R, \bar{\mu}_R , \mu_L\}$   are the Dirac spinors $\{\bar \tau, \bar \mu, \mu\}$ with the helicity operators, $(1\pm \gamma_5)/2$, respectively;
    $\sigma^{\alpha\beta}=\frac{i}{2}(\gamma^\alpha\gamma^\beta-\gamma^\beta\gamma^\alpha)$; 
    $D^\alpha=\partial^\alpha+ieA^\alpha$; $F^{\alpha \beta}=\partial^\alpha A^\beta-\partial^\beta A^\alpha$; $A^\alpha$ is the photon field; $e=-|e|$ is the electron charge; 
    $A_L$ and $A_R$ are the complex coefficients of interactions in which the intermediate photon has the left polarization and the right polarization, respectively; and
    $g_1,..., g_6$ are the complex coefficients of various 4 Fermi type interactions.

    According to Ref. \cite{Matsuzaki}, we can determine the observables,
\begin{align} \begin{split}\label{observables}
  a_\pm&=\frac{|g_1|^2}{16}\pm\frac{|g_2|^2}{16}+|g_3|^2\pm|g_4|^2
\\b_\pm&=|g_5|^2\pm|g_6|^2
\\c_\pm&=|eA_R|^2\pm|eA_L|^2
\\d_\pm&=-(Re[g_3eA_L^*]\pm Re[g_4eA_R^*])
\\e_\pm&=-(Re[g_6eA_R^*]\pm Re[g_5eA_L^*])
\\f_+&=-(Im[g_3eA_L^*]+Im[g_4eA_R^*])
\\g_+&=-(Im[g_6eA_R^*]+Im[g_5eA_L^*]).
\end{split} \end{align}
    $a_+$, $b_+$, $c_+$, $d_+$ and $e_+$ are determined from the final-state muon energy distribution and $a_-$, $b_-$, $c_-$, $d_-$, $e_-$, $f_+$ and $g_+$ are determined from the final-state angular distribution. 
    So, our first task is to determine $g_1,..., g_6$, $A_L$ and $A_R$ in this model.

    This paper is organized as follows.
    In Section \ref{SS2}, we derive the coupling constants $g_1,...,g_6,A_L$ and $A_R$ in this model.
    In Section \ref{SS3}, we give four scenarios and study the features of each scenario. 
    In Section \ref{SS4}, we discuss the difference between the minimal supersymmetric SM (MSSM) and Type-I\!I\!I 2HDM. 
    In Section \ref{SS5}, we  the summary and discussion.

\section{Effective Coupling Constants in the Model}\label{SS2}

    In Type-I\!I\!I 2HDM, $\tau\to3\mu$ decay can be written in tree level.
    However, we also consider the one-loop radiative diagrams since the resonance effect enhances the contribution as follows.
    These diagrams contain the photon propagator, which is proportional to $1/q^2$, where $q$ is the propagating momentum.
    The minimum value of $q^2$ is $4m_\mu^2$.
    This is realized when the pair created muon anti-muon have the same momentum.  
    On the other hand, in the Higgs mediated diagrams, this part is replaced by the Higgs mass squared.

\subsection{Four Fermi Diagrams}

\begin{figure}[htbp]
\begin{tabular}{cc}
\begin{minipage}{0.5\hsize}
\begin{center}
   \includegraphics[keepaspectratio=true,height=28mm]{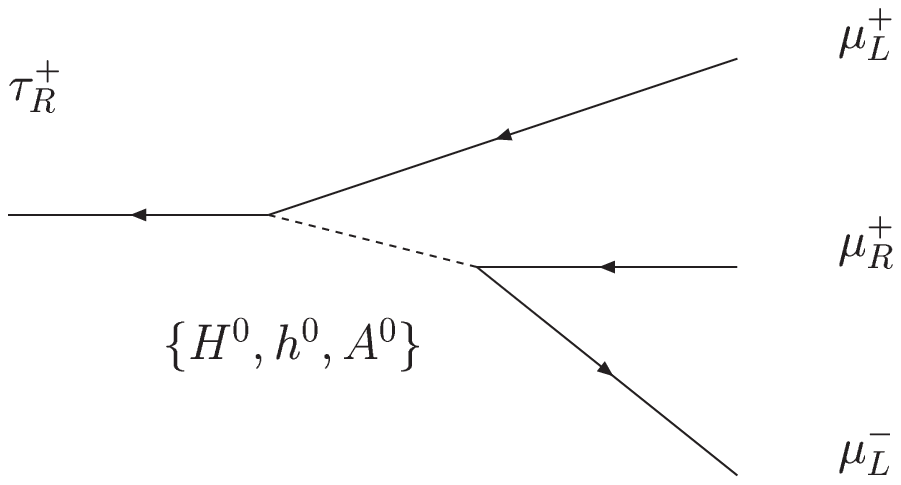}
  \caption{The tree-level diagram, which corresponds to the coupling constant $g_1$.}
\label{fig:1-1}
\end{center}
\end{minipage}
\ \ \ \ \ 
\begin{minipage}{0.5\hsize}
\begin{center}
   \includegraphics[keepaspectratio=true,height=28mm]{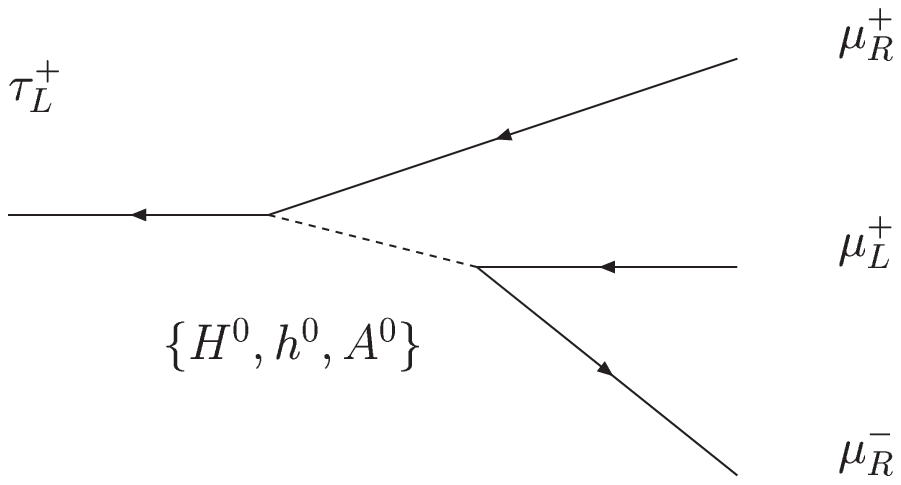}  \caption{The tree-level diagram, which corresponds to the coupling constant $g_2$.}
\label{fig:1-2}
\end{center}
\end{minipage}
\end{tabular}
\end{figure} 
\begin{figure}[htbp]
\begin{tabular}{cc}
\begin{minipage}{0.5\hsize}
\begin{center}
   \includegraphics[keepaspectratio=true,height=28mm]{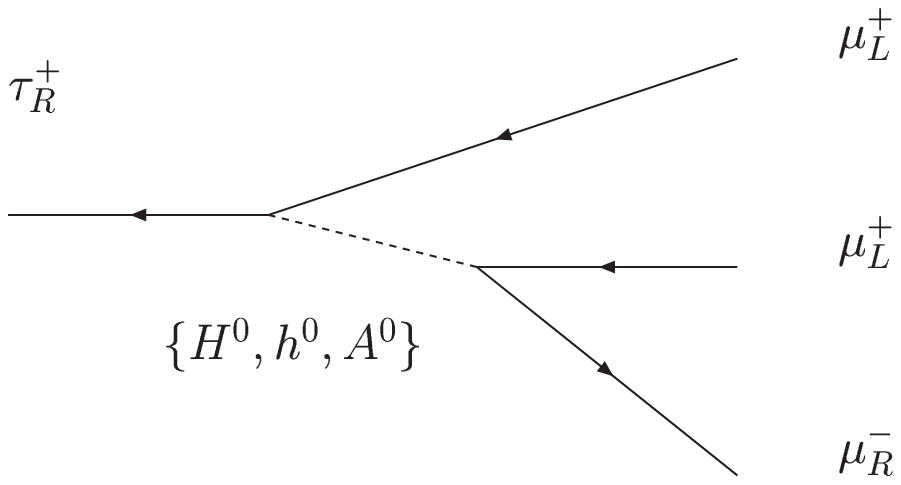}
  \caption{The tree-level diagram, which corresponds to the coupling constant $2g_5$.}
\label{fig:1-5}
\end{center}
\end{minipage}
\ \ \ \ \ 
\begin{minipage}{0.5\hsize}
\begin{center}
   \includegraphics[keepaspectratio=true,height=28mm]{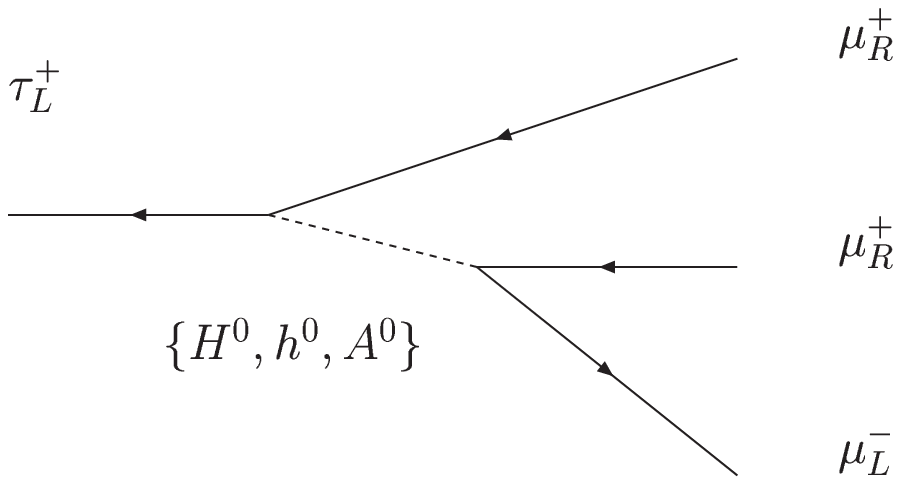}  \caption{The tree-level diagram, which corresponds to the coupling constant $2g_6$.}
\label{fig:1-6}
\end{center}
\end{minipage}
\end{tabular}
\end{figure} 
    Here, we calculate the tree-level diagrams which contain the neutral Higgs bosons in the intermediate state and are written as Figs. \ref{fig:1-1}, \ref{fig:1-2}, \ref{fig:1-5} and \ref{fig:1-6}.
    Comparing these diagrams and the general Lagrangian (\ref{Lag}), the effective coupling constants $g_1,g_2,g_5$ and $g_6$ contain 
\begin{align} \begin{split}
g_1 
&\supset
\frac{1}{- 2\sqrt{2}G_F}\biggl( J_{23}^{H^0*} J_{22}^{H^0*} \frac{1}{M_{H^0}^2}+J_{23}^{h^0*} J_{22}^{h^0*} \frac{1}{M_{h^0}^2}-J_{23}^{A^0*} J_{22}^{A^0*} \frac{1}{M_{A^0}^2}\biggr)
\\g_2
& \supset
\frac{1}{- 2\sqrt{2}G_F}\biggl( J_{32}^{H^0} J_{22}^{H^0}  \frac{1}{M_{H^0}^2}+J_{32}^{h^0} J_{22}^{h^0}  \frac{1}{M_{h^0}^2}-J_{32}^{A^0} J_{22}^{A^0}  \frac{1}{M_{A^0}^2}\biggr)
\\g_5 
&\supset 
\frac{1}{- 4\sqrt{2}G_F}\biggl( J_{23}^{H^0*} J_{22}^{H^0}  \frac{1}{M_{H^0}^2}+ J_{23}^{h^0*} J_{22}^{h^0}  \frac{1}{M_{h^0}^2}+ J_{23}^{A^0*} J_{22}^{A^0}  \frac{1}{M_{A^0}^2}\biggr)
\\g_6
& \supset
\frac{1}{- 4\sqrt{2}G_F}\biggl(   J_{32}^{H^0} J_{22}^{H^0*} \frac{1}{M_{H^0}^2}+ J_{32}^{h^0} J_{22}^{h^0*} \frac{1}{M_{h^0}^2}+ J_{32}^{A^0} J_{22}^{A^0*} \frac{1}{M_{A^0}^2}\biggr),
\end{split} \end{align}
where 
\begin{align} \begin{split}
J_{ij}^{H^0} 
&=
-\frac{g m_i \delta_{ij}}{2\sin\beta M_W}\sin\alpha  +\frac{ \eta_{ij}^E}{\sqrt{2}\sin\beta}\sin(\alpha-\beta) 
\\
J_{ij}^{h^0} &=
-\frac{g m_i \delta_{ij}}{2\sin\beta M_W}\cos\alpha  +\frac{ \eta_{ij}^E}{\sqrt{2}\sin\beta}\cos(\alpha-\beta) 
\\
J_{ij}^{A^0} &=
\frac{g m_i \delta_{ij}}{2\sin\beta M_W}\cos\beta  +\frac{\eta_{ij}^E}{\sqrt{2}\sin\beta}
\end{split} \end{align}
    are the effective coupling constants, where $g$ is the $\mathrm{SU(2)}_L$ gauge coupling constant;
    $m_i$ are the $i$-th family charged lepton masses;  
    $\alpha$ is mixing angle between neutral CP even Higgses;
    $\beta$ is defined as $\tan\beta=v_2/v_1$, where $v_1$ and $v_2$ are the vacuum expectation values of down and up-sector Higgses, respectively;
    $M_W$ is the weak boson mass;
    $H^0$ and $h^0$ are the heavier and lighter CP even neutral Higgses, respectively;
    $A^0$ is the CP odd neutral Higgs;
    $M_{H^0}$, $M_{h^0}$ and $M_{A^0}$ are $H^0$, $h^0$ and $A^0$ masses, respectively;
    $\eta_{ij}^E$ are non-diagonal leptonic Yukawa couplings of Type-I\!I\!I 2HDM.   
    Here, we note that $g_5$ and $g_6$ are generated via Fierz transformation.
    They are the same as $g_1$ and $g_2$ in their FCNC parts and flipping the helicity in their flavor conserving neutral current part, respectively.

\subsection{Radiative diagrams}\label{SS2.2}
    We consider the radiative one-loop diagrams since the resonance effect enhances its contribution as mentioned in the first of this section.
    
    In general, the anti-fermion to anti-fermion and imaginary photon FCNC amplitude which momenta are $p$, $p-q$ and $q$, respectively, is written as 
\begin{align} \begin{split}
\bar{v}(p)i\biggl[
 q^2\gamma^\mu(C_1 P_R+C_2 P_L)
+ q^\mu (C_3 P_R+C_4 P_L)
+i\sigma^{\nu\mu}q_\nu (C_5 P_R+C_6 P_L)
\biggr]v(p-q),
\end{split} \end{align}
    where $C_1,...,C_6$ are the complex coefficients which should be given by the model;
    $P_R,P_L\equiv (1\pm\gamma^5)/2$; 
    $\bar{v}(p)$ and $v(p-q)$ are the spinors of initial and final anti-fermion, respectively.

    Considering $\tau\to3\mu$ decay, $q^\mu (C_3 P_R+C_4 P_L)$ term will vanish when we multiply the electromagnetic current $-ie\bar{\mu} \gamma_\mu \mu$ and use the Dirac equation.
    By the same prescription, $ q^2\gamma^\mu(C_1 P_R+C_2 P_L)$ term becomes
\begin{align} \begin{split}
& \bar{v}(p)i\biggl[q^2\gamma^\mu(C_1 P_R+C_2 P_L)\biggr]v(p-q)\frac{-ig_{\mu\nu}}{q^2}
 \bar{u}(p_3) (-ie\gamma^\nu)( P_R+ P_L) v(p_2),
  \end{split} \end{align}
    where $p_2$ and $p_3$ are the final-state fermion and anti-fermion momenta, respectively.
    So, $g_3,...,g_6$ contain $C_1$ and $C_2$ as
\begin{align} \begin{split}
   g_3\supset \frac{eC_1}{ 2\sqrt{2}G_F}
,&\ \  g_4\supset \frac{eC_2}{ 2\sqrt{2}G_F}
\\ g_5\supset \frac{eC_1}{ 2\sqrt{2}G_F} 
,&\ \  g_6\supset \frac{eC_2}{ 2\sqrt{2}G_F}.
\end{split} \end{align}
    Also, $i\sigma^{\nu\mu}q_\nu (C_5 P_R+C_6 P_L)$ should be compared with $A_L$ and $A_R$. 
    They contain $C_5$ and $C_6$ as
\begin{align} \begin{split}
  A_L \supset \frac{C_5}{-4\sqrt{2}G_F m_\tau},
\ \   A_R \supset \frac{C_6}{-4\sqrt{2}G_F m_\tau}.
\end{split} \end{align}

\begin{figure}[t]
\begin{tabular}{cc}
\begin{minipage}{0.5\hsize}
\begin{center}
   \includegraphics[keepaspectratio=true,height=28mm]{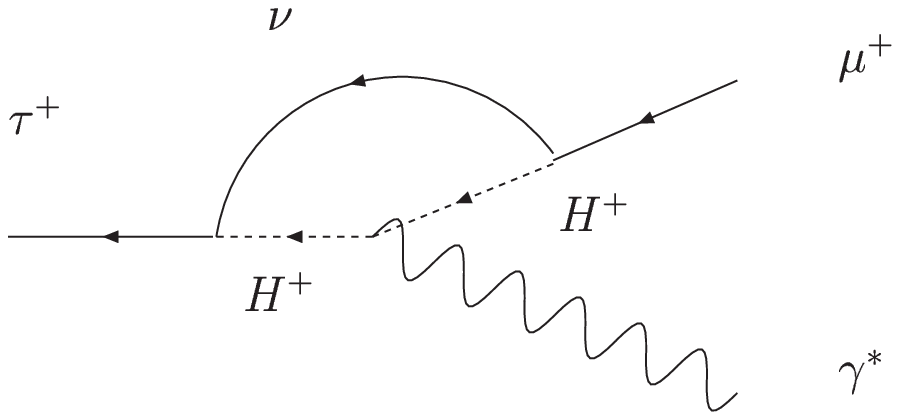}
  \caption{The radiative one-loop diagram which have charged Higgs bosons and neutrinos in the loop.}
\label{fig:rad H+}
\end{center}
\end{minipage}
\ \ \ \ \ \ 
\begin{minipage}{0.5\hsize}
\begin{center}
   \includegraphics[keepaspectratio=true,height=28mm]{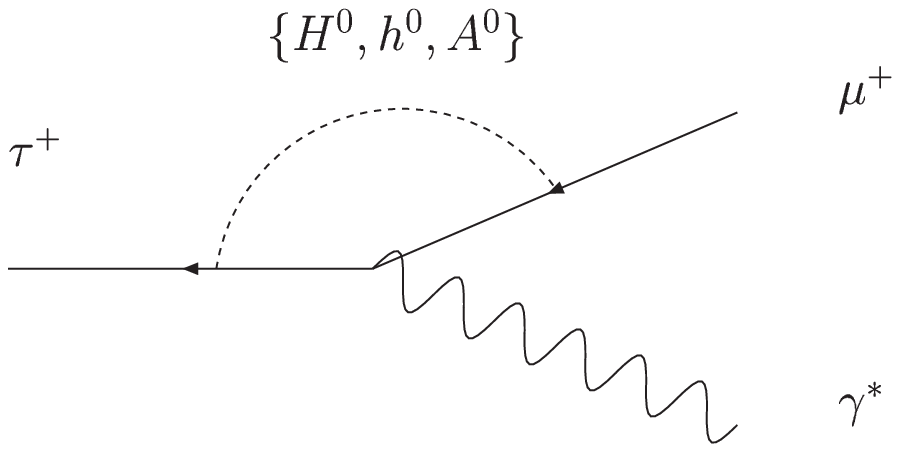} 
    \caption{The radiative one-loop diagram which have neutral Higgs bosons and charged leptons in the loop.
    The dashed line represents $H^0$, $h^0$ and $A^0$ propagator.}
\label{fig:rad H0}
\end{center}
\end{minipage}
\end{tabular}
\end{figure} 

    In Type-I\!I\!I 2HDM, these diagrams are written in Figs. \ref{fig:rad H+} and \ref{fig:rad H0}.
    The coefficients $C_1,C_2,C_5$ and $C_6$ are calculated in Appendix \ref{Appendix calc}, and the explicit form of effective coupling constants $g_1,...,g_6$, $A_L$ and $A_R$ are written in Appendix \ref{coupling constants}.

\section{Scenarios}\label{SS3}

    Here, we suppose four possible scenarios in type-I\!I\!I 2HDM.
    They are as follows:
\begin{enumerate}
  \item All Higgs bosons except for $h^0$ are very heavy and decoupled.
  \item All Higgs bosons except for $A^0$ are very heavy and decoupled.
  \item $H^\pm$ and $A^0$ are very heavy and decoupled, where $H^\pm$ are the charged Higgs bosons which have $\pm1$ electromagnetic charges, respectively. 
    $h^0$ and $H^0$ have the masses $M_{h^0}=98$GeV and $M_{H^0}=115$GeV, respectively. 
    $ZZh^0$ coupling is too small to be discovered in LEP experiment. 
  \item All Higgs bosons except for $H^\pm$ are very heavy and decoupled.
\end{enumerate}
    Each scenario suggests different final-state distributions.
    Comparing them to the experimental result, we can test the Type-I\!I\!I 2HDM.
    
\subsection{Scenario 1: only $h^0$ is light}\label{s3.1}

    If we discover only one neutral Higgs boson in LHC, there may be no signal of new physics in LHC.
    However, if the LFV is discovered, we have to reconsider whether the Higgs sector is the SM one or not.

    In this scenario, we set $M_{H^0},M_{A^0},M_{H^\pm}\gg M_{h^0}$, where $M_{H^\pm}$ is $H^\pm$ mass. 
    This situation becomes important when we discover only one neutral Higgs boson in LHC.
    
    According to the above mass relation and Appendix \ref{coupling constants}, we find the relations
\begin{align} \begin{split}\label{1-3relations}
g_5\simeq\frac{g_1}{2}\frac{J_{22}^{h^0}}{J_{22}^{h^0*}}+g_3,
\ \ \ \ \
g_6\simeq\frac{g_2}{2}\frac{J_{22}^{h^0*}}{J_{22}^{h^0}}+g_4.
\end{split} \end{align}
    Substituting Eq. (\ref{observables}) for $g_3$, $g_4$, $g_5$ and $g_6$ in these relations, and then solving about $|g_1|^2$ and $|g_2|^2$, respectively,  %using the fact that $|J_{22}^{h^0}/J_{22}^{h^0*}|=1$,
     %$|g_1|^2$ and $|g_2|^2$ 
     they are written by the observables as follows:
\begin{align} \begin{split}
|g_1|^2&\simeq\frac{8}{25} \left(5  (a_++a_-)+3 (b_++b_-)
-2(e_+-e_-)\frac{5(d_++d_-)-(e_+-e_-)}{c_+-c_-}\right)
\\&\pm
\frac{16}{25} \sqrt{ (b_++b_-)-\frac{(e_+-e_-)^2}{c_+-c_-}}
\sqrt{4 \bigl( 5  (a_++a_-)-(b_++b_-)\bigr)
-\frac{\bigl(5 (d_++d_-)-(e_+-e_-)\bigr)^2}{c_+-c_-}},
\\|g_2|^2&\simeq\frac{8}{25} \left(5  (a_+-a_-)+3 (b_+-b_-)
-2(e_++e_-)\frac{5(d_+-d_-)-(e_++e_-)}{c_++c_-}\right)
\\&\pm
\frac{16}{25} \sqrt{ (b_+-b_-)-\frac{(e_++e_-)^2}{c_++c_-}}
\sqrt{4 \bigl( 5  (a_+-a_-)-(b_+-b_-)\bigr)
-\frac{\bigl(5 (d_+-d_-)-(e_++e_-)\bigr)^2}{c_++c_-}}.
\end{split} \end{align}
    These equations have the discrete ambiguities in their later terms.
    These correspond to the sign of imaginary part of $g_5eA_L^*$ and $g_6eA_R^*$, respectively.
    Moreover, using these quantities, we determine    
\begin{align} \begin{split}
|g_3|^2\simeq\frac{a_++a_-}{2}-\frac{|g_1|^2}{16},\ \ \ \ \ \ 
|g_4|^2\simeq\frac{a_+-a_-}{2}-\frac{|g_2|^2}{16},
\end{split} \end{align}
and
\begin{align} \begin{split}
Arg[g_3eA_L^*]&
\simeq\arccos\Bigl[-\frac{d_++d_-}{\sqrt{2}|g_3|\sqrt{c_+-c_-}}\Bigr]
\\Arg[g_4eA_R^*]&
\simeq\arccos\Bigl[-\frac{d_+-d_-}{\sqrt{2}|g_4|\sqrt{c_++c_-}}\Bigr]
\\Arg[g_1\frac{J_{22}^{h^0}}{J_{22}^{h^0*}}g_3*]&
\simeq\arccos\Bigl[\frac{8(b_++b_-)-8(a_++a_-)-3|g_1|^2}{16|g_1||g_3|}\Bigr]
\\Arg[g_2\frac{J_{22}^{h^0*}}{J_{22}^{h^0}}g_4*]&
\simeq\arccos\Bigl[\frac{8(b_+-b_-)-(a_+-a_-)-3|g_2|^2}{16|g_2||g_4|}\Bigr]
\\Arg[g_3g_5^*]&
\simeq\arccos\Bigl[\frac{(b_++b_-)+10|g_3|^2-4(a_++a_-)}{2\sqrt{2}|g_3|\sqrt{b_++b_-}}\Bigr]
\\Arg[g_4g_6^*]&
\simeq\arccos\Bigl[\frac{(b_+-b_-)+10|g_4|^2-4(a_+-a_-)}{2\sqrt{2}|g_4|\sqrt{b_+-b_-}}\Bigr].
\end{split} \end{align}

\subsubsection{Situation division}
    This scenario contains three types of couplings as follows:
    
\begin{enumerate}
  \item $J_{31}^{h^0}J_{12}^{h^0}$, $J_{31}^{h^0}J_{21}^{h^0*}$, $J_{13}^{h^0*}J_{12}^{h^0}$, $J_{13}^{h^0*}J_{12}^{h^0*},$
  \item $J_{32}^{h^0}J_{22}^{h^0}$, $J_{32}^{h^0}J_{22}^{h^0*}$, $J_{23}^{h^0*}J_{22}^{h^0}$, $J_{23}^{h^0*}J_{22}^{h^0*},$
  \item $J_{33}^{h^0}J_{32}^{h^0}$, $J_{33}^{h^0}J_{23}^{h^0*}$, $J_{33}^{h^0*}J_{32}^{h^0}$, $J_{33}^{h^0*}J_{23}^{h^0*}.$
\end{enumerate}
    These are divided by the family of intermediate leptons. 
    The tree-level contribution is contained in the second-type coupling.
    However, we cannot neglect other contributions since each $J_{ij}^{h^0}$ is only a parameter of the model.

    To determine which contributions are dominant, we first divide in four cases using the observables which are defined in Eq. (\ref{observables}) and the relations (\ref{1-3relations}) as
\begin{align} \begin{split}
\frac{a_+}{b_+}&\simeq
\frac{|g_1|^2+|g_2|^2+16(|g_3|^2+|g_4|^2)}
{4(|g_1|^2+|g_2|^2)+16(|g_3|^2+|g_4|^2)+16Re[g_1\frac{J_{22}^{h^0}}{J_{22}^{h^0*}}g_3^*
+g_2\frac{J_{22}^{h^0*}}{J_{22}^{h^0}}g_4^*]},
\end{split} \end{align}
    $d_+/c_+$ and $c_+/a_+$.

\subsubsection*{case 1; $a_+/b_+\simeq 1/4$}

    When $a_+/b_+\simeq 1/4$, $|g_1|^2+|g_2|^2$ is dominant and this means that the tree-level contribution i.e. second-type coupling works.
    So, even if all $|J_{ij}^{h^0}|$ are the same order, the observable $a_+/b_+$ suggests this case.
    We note here that $c_+/a_+\ll1$ in this case.

\subsubsection*{case 2; $a_+/b_+\simeq 1$ and $c_+/a_+=\mathcal{O}(1) $}

    When $a_+/b_+\simeq 1$, then $|g_3|^2+|g_4|^2\gg|g_1|^2+|g_2|^2$.
    In this case, the second-type coupling $|J_{23}^{h^0*} J_{22}^{h^0*}|^2+ |J_{32}^{h^0} J_{22}^{h^0}|^2$ is highly suppressed compared with other two types of couplings.
    From the equations in Appendix \ref{coupling constants}, $|J_{23}^{h^0*} J_{22}^{h^0*}|^2+ |J_{32}^{h^0} J_{22}^{h^0}|^2$ is highly suppressed as
\begin{align} \begin{split}
\mathcal{O}(10^{-5})
&\gg \frac{|J_{23}^{h^0*} J_{22}^{h^0*}|^2+ |J_{32}^{h^0} J_{22}^{h^0}|
^2}
{|J_{13}^{h^0*}J_{12}^{h^0}|^2+|J_{31}^{h^0}J_{21}^{h^0*}|^2+|J_{33}^{h^0*}J_{32}^{h^0}|^2+|J_{33}^{h^0}J_{23}^{h^0*}|^2}.
\end{split} \end{align}
    The left hand side in above relation is small since the tree-level diagrams affect $|g_1|^2+|g_2|^2$, while only one-loop diagrams affect $|g_3|^2+|g_4|^2$.
    The radiative one-loop currents are dominant to the tree-level FCNC.
    So, $\tau\to\mu\gamma$ may be discovered earlier than $\tau\to3\mu$.
    However, $\tau\to3\mu$ mode has the advantage that we can determine which of the third-type couplings and the first-type couplings are dominant using the observable $c_+/a_+$. 
    If the third-type couplings are dominant, we have an observable relation from Appendix \ref{coupling constants} as,
\begin{align} \begin{split}\label{c/a-3}
\frac{c_+}{a_+}&
\simeq\frac{9}{4}
+\frac{27}{8}\left(1-\frac{ Re[J_{33}^{h^0 2}]}{|J_{33}^{h^0}|^2}\right)
 \frac{3+2\log[\frac{m_\tau^2}{M_{h^0}^2}]}
 {(4+3\log[\frac{m_\tau^2}{M_{h^0}^2}]) ^2}.
 \end{split} \end{align}
    When we set $M_{h^0}=115$GeV,
\begin{align} \begin{split}
\frac{9}{4}-0.21\siml\frac{c_+}{a_+}&\siml\frac{9}{4}.
 \end{split} \end{align}

    In this case, there is one more simple relation,
\begin{align} \begin{split}
\frac{d_+}{e_+}\simeq1.
\end{split} \end{align}
    We can confirm $|g_3|^2+|g_4|^2$ dominance using these relations.

\subsubsection*{case 3; $a_+/b_+\simeq 1$ and $c_+/a_+=\mathcal{O}(10^{-4}) $}

    If $a_+/b_+\simeq 1$ as the case 2, however the first-type couplings are dominant and no tuning, 
\begin{align} \begin{split}
\frac{c_+}{a_+}
\simeq
 \frac{9}{ 16}\left(-1+3\log[\frac{q^2}{M_{h^0}^2}]\right)^{-2}
 \simeq0.00034,
\end{split} \end{align}
    where $M_{h^0}=115$GeV.
    These two cases have different order of $c_+/a_+$ values.
    This fact is an advantage to determine the current structure.

    In this case, we note 
\begin{align} \begin{split}
\frac{d_+}{e_+}\simeq1
\end{split} \end{align}
    as in case 2.

\subsubsection*{case 4; $a_+/b_+\not\simeq 1, 1/4$}
    If $a_+/b_+\not\simeq1$ or $1/4$, this means $|g_1|^2+|g_2|^2 \sim |g_3|^2+|g_4|^2$.
    This case is similar to the case 2 and the case 3 with no tuning, and
\begin{align} \begin{split}
\mathcal{O}(10^{-5})
&\sim \frac{|J_{23}^{h^0*} J_{22}^{h^0*}|^2+ |J_{32}^{h^0} J_{22}^{h^0}|
^2}
{|J_{13}^{h^0*}J_{12}^{h^0}|^2+|J_{31}^{h^0}J_{21}^{h^0*}|^2+|J_{33}^{h^0*}J_{32}^{h^0}|^2+|J_{33}^{h^0}J_{23}^{h^0*}|^2}.
\end{split} \end{align}

\subsubsection{Bounds from Total Branching Ratio}

    We give the upper limits on the effective couplings in each case studied above.
    Supposing parity and CP conservation for simplicity, the total branching ratio is written as \cite{Matsuzaki}
\begin{align} \begin{split}
\mathrm{Br}\bigl(\tau^+\to \mu_1 \mu_2 \mu_3\bigr) 
  &= \mathrm{Br}(\tau\to \mu \nu\bar\nu)
  \Biggl[ 2a_+ + b_+
+16d_+ +8e_+
+\frac{8}{3}\Bigl(24\log\bigl[\frac{3}{4\delta}\bigr]-13\Bigr)c_+ \Biggr],
\end{split} \end{align} 
    where $\delta=2m_\mu/m_\tau$. 

%%%%%%%%%%%%%%%%%%%%%%%%%%%%%%%%%%%%%%%%%%%%%%%%%%%%%%%%%%%%%%%%%%%%%%%55

\begin{figure}[t]
\begin{center}
   \includegraphics[keepaspectratio=true,height=58mm]{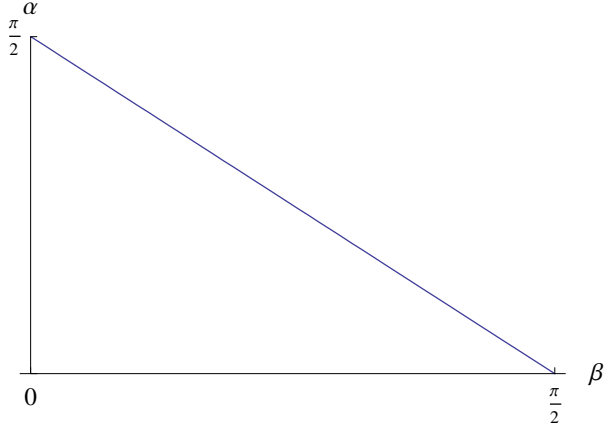}
  \caption{Relation between $\alpha$ and $\beta$ in scenario 1.}
\label{fig:ab-1-1}
\end{center}
\end{figure}

    To reduce the free parameter of the model, we consider the relation between $\alpha$ and $\beta$.
    In this scenario, according to the Appendix \ref{Appendix Higgs}, we set $M_{11}M_{22}=M_{12}^2$ since $M_{H^0}\gg M_{h^0}$.
    Here, we set $\lambda_6$ is large enough since $ M_{A^0}^2=\lambda_6(v_1^2+v_2^2)$ is decoupled, and for CP conservation of Higgs potential, $\lambda_6=\lambda_5$. 
    So, $\lambda_6=\lambda_5 \gg\lambda_1,...,\lambda_3$ and
\begin{align} \begin{split}
  M_{11}&=v_2^2\lambda_5
\\M_{22}&=v_1^2\lambda_5
\\M_{12}&=M_{21}=\lambda_5 v_1 v_2.
\end{split} \end{align}
    Here, we set the $\alpha$ region as follows:
\begin{align} \begin{split}
 M_{A^0}^2>0           \Rightarrow
 \lambda_6=\lambda_5>0 \Rightarrow 
 M_{12}>0              \Rightarrow 
 \sin2\alpha>0         \Rightarrow 
 0< \alpha<\frac{\pi}{2}.
\end{split} \end{align}
    Using these conditions, we have the relations from Appendix \ref{Appendix Higgs} as,
\begin{align} \begin{split}
\tan \alpha &= \sqrt{\frac{M_{22}}{M_{11}}}.
\\\frac{1}{\tan^2\alpha}=\frac{M_{11}}{M_{22}}
&\simeq\frac{v_2^2}{v_1^2}=\tan^2\beta,
\\\cos\alpha
&=
\sin\beta
\\
\cos(\alpha-\beta)
&=
2\sin\beta\cos\beta.
\end{split} \end{align}
    This is explained in Fig. \ref{fig:ab-1-1}

\subsubsection*{Second-type coupling: $J_{23}^{h^0}  J_{22}^{h^0} $}

    We here consider the case that second-type coupling i.e. $J_{23}^{h^0}  J_{22}^{h^0} $ is dominant, which case corresponds to the case 1 in above. 
    In this case, only $g_1=g_2=2g_5=2g_6$ are considerable and the branching ratio is given as 
\begin{align} \begin{split}
\mathrm{Br}\bigl(\tau^+\to \mu_1 \mu_2 \mu_3\bigr)
&\simeq 4.8 \times (\eta_{23}^E \eta_{22}^E )^2
\cos^4\beta\Bigl(\frac{100\mathrm{GeV}}{M_{h^0}}\Bigr)^4.
\end{split} \end{align}
    If we set $M_{h^0}=115$GeV, $\cos\beta=1/\sqrt{2}$ and $ \mathrm{Br}(\tau\to 3\mu)<3.2\times 10^{-8}$ from \cite{Bell Br}, then the upper limit is written as
\begin{align} \begin{split}
|\eta_{23}^E \eta_{22}^E|<0.00022.
\end{split} \end{align}

\subsubsection*{First-type coupling: $J_{13}^{h^0}  J_{12}^{h^0} $}
    We here consider the case that the first-type coupling i.e. $J_{13}^{h^0}  J_{12}^{h^0} $ is dominant, which case corresponds to the case 3 in above.
    Then $g_1,g_2\simeq0$, $g_5=g_6=g_3= g_4$, $eA_L = eA_R $ and the branching ratio is written as 
\begin{align} \begin{split}
&\mathrm{Br}\bigl(\tau^+\to \mu_1 \mu_2 \mu_3\bigr) 
\\&\simeq0.14\times10^{-5}\times  \Bigl[0.73+ \bigl(-12.3+  \log[\Bigl(\frac{100GeV}{M_{h^0}}\Bigr)^2]\bigr)^2\Bigr]
 \Bigl(\frac{100GeV}{ M_{h^0}}\Bigr)^4 \cos^4\beta
(\eta_{13}^E \eta_{12}^E)^2. 
\end{split} \end{align} 
    If we set $M_{h^0}=115$GeV, $\cos\beta=1/\sqrt{2}$ and $ \mathrm{Br}(\tau\to 3\mu)<3.2\times 10^{-8}$, we give an upper limit as
\begin{align} \begin{split}
|\eta_{13}^E \eta_{12}^E|\le0.032.
\end{split} \end{align}

\subsubsection*{Third-type coupling: $J_{33}^{h^0}  J_{32}^{h^0} $}
    We here consider the case that the third-type coupling $J_{33}^{h^0}  J_{32}^{h^0} $ is dominant, which case corresponds to the case 2 in above.
    Then $g_1,g_2\simeq0$, $g_5=g_6=g_3$ and
\begin{align} \begin{split}
&\mathrm{Br}\bigl(\tau^+\to \mu_1 \mu_2 \mu_3\bigr) 
\\&\simeq
1.1\times10^{-4}   (\cos^2\beta(-\frac{0.0051}{\cos\beta} +\eta_{33}^E)\eta_{23}^E)^2 
(6.7- \log[\Bigl(\frac{100\mathrm{GeV}}{M_{h^0}}\Bigr)^2])^2\Bigl(\frac{100\mathrm{GeV}}{M_{h^0}}\Bigr)^4.
\end{split} \end{align}
    If we set $M_{h^0}=115$GeV, $\cos\beta=1/\sqrt{2}$ and $ \mathrm{Br}(\tau\to 3\mu)<3.2\times 10^{-8}$, we give an upper limit as
\begin{align} \begin{split}
|(-0.0072+\eta_{33}^E)\eta_{23}^E | \le0.0065.
\end{split} \end{align}

\subsection{Scenario 2: only $A^0$ is light}
    Here, we study the scenario 2, in which all Higgs bosons except for $A^0$ are decoupled.
    In this case, $g_5$ and $g_6$ are written from Appendix \ref{coupling constants} as
\begin{align} \begin{split}
 g_5\simeq -\frac{g_1}{2}\frac{J_{22}^{A^0}}{J_{22}^{A^0*}}+g_3,
\ \ \ \ \
g_6 \simeq -\frac{g_2}{2}\frac{J_{22}^{A^0*}}{J_{22}^{A^0}}+g_4.
\end{split} \end{align}
    The observable difference from the scenario 1 presents in $a_+/c_+$ when $J_{33}^{A^0}J_{32}^{A^0}$ is dominant and CP violation in the Lagrangian is small.
    When  $J_{13}^{A^0} {}^* J_{12}^{A^0}$ or its hermit conjugate (h.c.) is dominant, the difference from  $J_{13}^{h^0} {}^* J_{12}^{h^0}$ or its h.c. dominant situation is suppressed by $m_e/m_\tau$.
    When $J_{23}^{A^0}{}^*J_{22}^{A^0}$ or its h.c. is dominant, the difference from  $J_{23}^{h^0} {}^* J_{22}^{h^0}$ or its h.c. dominant situation is suppressed by the fine-structure constant $\alpha_{QED}=e^2/(4\pi)$.

    In scenario 1, only $h^0$ is light, if CP symmetry of Lagrangian is conserved, $Re[(J^{h^0}_{33})^2]=|J_{33}^{h^0*}|^2$ and
\begin{align} \begin{split}
\frac{c_+}{a_+}&\simeq
\frac{9}{4},
\end{split} \end{align}
    from Eq. (\ref{c/a-3}).
    On the other hand, in this scenario, only $A^0$ is light, if CP symmetry of Lagrangian is conserved,
\begin{align} \begin{split}
\frac{c_+}{a_+}&\simeq
\frac{9}{4}
+\frac{27}{8}\left(1+\frac{ Re[J_{33}^{A^0 2}]}{|J_{33}^{A^0}|^2}\right)
 \frac{3+2\log[\frac{m_\tau^2}{M_{A^0}^2}]}
 {(4+3\log[\frac{m_\tau^2}{M_{A^0}^2}]) ^2}.
%
%\frac{9}{4}
%\Biggl[1+\frac{3}{2}\frac{(3+2\log[\frac{m_\tau^2}{M_{A^0}^2}])}
%{(4+3\log[\frac{m_\tau^2}{M_{A^0}^2}])^2 }(1+ \frac{Re[(J^{A^0}_{33})^2]
%}{|J_{33}^{A^0*}|^2} )\Biggr]
\simeq \frac{9}{4}-0.21,
\end{split} \end{align}
    where we set $M_{A^0}=115$GeV.
    This value is stable even if $M_{A^0}$ is varied intensely since it is in logarithmic function.

\subsection{Scenario 3: only $H^0$ is seen}

    In this scenario, we can detect only $H^0$.
    We suppose that $h^0$, which is lighter than $H^0$, could not be detected in LEP since the $ZZh^0$ coupling in 2HDM is too weak.
    $H^\pm$ and $A^0$ are also decoupled since they are very heavy.
    The LEP experiment suggests a possibility of $M_{H^0}=115$GeV and $M_{h^0}=98$GeV \cite{LEP}.
    In this scenario, we set from Appendix \ref{Appendix Higgs} as,
\begin{align} \begin{split}
\frac{M_{11}}{M_{22}}&\simeq\tan^2\beta
\\
\lambda_5=\lambda_6 &\gg \lambda_{1,2,3}
\\
\cos2\alpha&\simeq-\frac{M_{H^0}+M_{h^0}}{M_{H^0}-M_{h^0}}\cos2\beta=-\frac{213}{17}\cos2\beta.
\end{split} \end{align}

    This relation between $\alpha$ and $\beta$ suggests that $\beta\simeq \pi/4$ in all $\alpha$ range. 
    So, the effective coupling constant $J_{ij}^{H^0}$ is written as
\begin{align} \begin{split}
J_{ij}^{H^0}
\simeq
-\frac{g m_i \delta_{ij}}{\sqrt{2} M_W}\sin\alpha +\eta_{ij}^E\sin(\alpha-\frac{\pi}{4}).
\end{split} \end{align}

\begin{figure}[htbp]
\begin{center}
   \includegraphics[keepaspectratio=true,height=60mm]{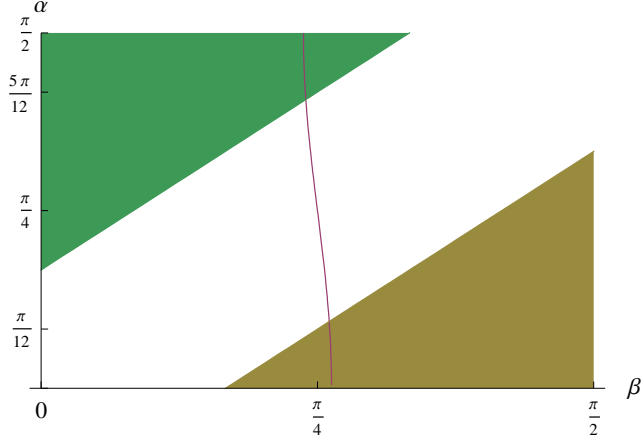}
  \caption{The allowed region of $\alpha$ and $\beta$ in scenario 3. 
            The upper left and the lower right are forbidden by the LEP constraint i.e. $\beta-\pi/6 <\alpha<\beta+\pi/6 $.
            If $M_{h^0}=98$GeV and $M_{H^0}=115$GeV, only on the curved line near $\beta=\pi/4$ is allowed. }
\label{fig:alpha-beta}
\end{center}
\end{figure}

    Also, from Refs. \cite{LEP} and \cite{Kim}, we set $|g_{ZZh^0}^{2HDM}|<|g_{ZZH}^{SM}|/2$, where $g_{ZZh^0}^{2HDM}$ and $g_{ZZH}^{SM}$ are the $ZZh^0$ coupling in Type-I\!I\!I 2HDM and $ZZ$ Higgs coupling in the SM, respectively.
    This suggests the allowed region $|\sin(\beta-\alpha)|<1/2$ since $ZZh^0$ term in 2HDM Lagrangian is $M_Z^2\sin(\beta-\alpha)ZZh^0/\sqrt{v_1^2+v_2^2}$, where $M_Z$ is the $Z$ boson mass.
    Using the fact $\beta\simeq \pi/4$, the allowed region of $\alpha$ becomes $\frac{\pi}{12}<\alpha<\frac{5\pi}{12}$.
    After all, the allowed region of $\alpha$ and $\beta$ in this scenario is explained as in Fig. \ref{fig:alpha-beta}.

    Comparing between Fig. \ref{fig:ab-1-1} and Fig. \ref{fig:alpha-beta}, this scenario has narrower allowed region than the scenario 1.
    So, if we can determine $\alpha$ or $\beta$, we can distinguish this scenario and scenario 1.

\subsection{Scenario 4: only $H^+$ is light}

    We consider all the Higgs bosons except for $H^+$ are decoupled.
    Here, we don't consider the parity and CP conservation.
    In this case, we have the relations from Appendix \ref{coupling constants} as,
\begin{align} \begin{split}
g_1 \simeq g_2& \simeq g_4 \simeq g_6\simeq0
\\g_3 \simeq g_5& \simeq \frac{4}{3}eA_L \simeq \frac{4}{3}\frac{m_\tau}{m_\mu}eA_R \simeq -\frac{\alpha_{QED}}{2\sqrt{2}G_F}
\frac{(J^{H^+\dagger} J^{H^+})_{32}}{9(4\pi) M_{H^+}^2},
\end{split} \end{align}
    and the observables become
\begin{align} \begin{split}
  c_+&\simeq -c_-\simeq \frac{9}{16}|g_3|^2
\\a_\pm& \simeq b_\pm \simeq |g_3|^2
\\d_\pm& \simeq \pm e_\pm \simeq -\frac{3}{4}|g_3|^2.
\end{split} \end{align}

\begin{align} \begin{split}
&\mathrm{Br}\bigl(\tau^+\to \mu_1 \mu_2 \mu_3\bigr) 
\\
&\simeq 5.3\times 10^{-8}\frac{1}{\sin^4\beta}\biggl(\frac{100GeV}{M_{H^+}}\biggr)^2|(\eta_{13}^{E*} \eta_{12}^E+\eta_{23}^{E*} \eta_{22}^E+(- 0.01\cos\beta +\eta_{33}^{E*})\eta_{32}^E)|^2
\end{split} \end{align}
    If we set $M_{H^\pm}=100$GeV, $\cos\beta=\sin\beta=1/\sqrt{2}$ and $ \mathrm{Br}(\tau\to 3\mu)<3.2\times 10^{-8}$, we give an upper limit as
\begin{align} \begin{split}
|(\eta_{13}^{E*} \eta_{12}^E+\eta_{23}^{E*} \eta_{22}^E+(- 0.0072  +\eta_{33}^{E*})\eta_{32}^E)|<0.39
\end{split} \end{align}

%%%%%%%%%%%%%%%%%%%%%%%%%%%%%%%%%%%%%%%%%%%%%%
%%%%%%%%%%%%%%%%%%%%%%%%%%%%%%%%%%%%%%%%%%%%%%
%%%%%%%%%%%%%%%%%%%%%%%%%%%%%%%%%%%%%%%%%%%%%%
%%%%%%%%%%%%%%%%%%%%%%%%%%%%%%%%%%%%%%%%%%%%%%

\section{Difference Between SUSY and Type-I\!I\!I 2HDM}\label{SS4}

\subsection{MSSM}
    In most MSSM models, the dominant contribution is from the radiative diagrams \cite{Okada}, \cite{MSSM2}.
    So, at first, the LFV event will be discovered in radiative mode e.g. $\tau\to\mu\gamma$.
    
    According to section \ref{SS2.2}, this feature appears in the effective coupling constants as
\begin{align} \begin{split}
g_1=g_2=0,
\ \ \ g_3=g_5,
\ \ \ g_4=g_6,
\end{split} \end{align}
    in $\tau\to3\mu$ mode. 
    So, the observables $a_\pm$, $b_\pm$, $d_\pm$, $e_\pm$, $f_+$ and $g_+$ have the relations,
\begin{align} \begin{split} \label{52}
        a_\pm=b_\pm,
\ \ \ \ d_\pm=\pm e_\pm,
\ \ \ \ f_+=g_+=0.
\end{split} \end{align}
    We note here that imaginary part becomes zero since $g_3$, $g_4$, $eA_L$ and $eA_R$ have the same complex phases.

     On the other hand, the tree-level diagram contributions in type-I\!I\!I 2HDM derives the characteristic relations, $g_1\not=0$ and $g_2\not=0$.
     Furthermore, in generally, tree-level heavy gauge boson in intermediate state derives the relation, $a_\pm\not=b_\pm$.
     In these case we can easily discriminate the difference between MSSM and 2HDM.

    Even if the experimental result suggests the relation (\ref{52}), it may still be Type-I\!I\!I 2HDM.
    In this case, another observable $a_+/c_+$ becomes important to distinguish them.
    If there is no tuning, this situation is realized when $J_{22}=0$ or $J_{23}^*=J_{32}=0$.
    So, if $a_+/c_+\sim 9/4$ or $\sim 0.0034$ as derived in section \ref{s3.1}, 2HDM is strongly supported. 
    On the other hand, if $a_+/c_+$ has other values, then the MSSM is supported. 

\subsection{Babu-Kolda Model}
    In Babu-Kolda model (BKM) \cite{Babu}, MSSM neutral higgses propagate the intermediate state of the  $\tau\to3\mu$ decay.
    If we consider the situation, $Br(\tau\to\mu\gamma) \hspace{0.3em}\raisebox{0.4ex}{$<$}\hspace{-0.75em}\raisebox{-.7ex}{$\sim$}\hspace{0.3em}Br(\tau\to3\mu)$, the radiative diagrams affect little in $\tau\to3\mu$ decay.
    So, we now neglect it.
    In this model, $g_1$ and $g_5$ are written as
\begin{align} \begin{split}
g_1&=
\frac{m_\tau m_\mu \kappa_{32}}{2\cos^3\beta}
\Bigl(\frac{\cos(\alpha-\beta)\sin\alpha}{M_h^2}+\frac{\sin(\alpha-\beta)\cos\alpha}{M_H^2}-\frac{\sin\beta}{M_A^2}\Bigr)
\\g_5&=\frac{1}{2}
\frac{m_\tau m_\mu \kappa_{32}}{2\cos^3\beta}
\Bigl(\frac{\cos(\alpha-\beta)\sin\alpha}{M_h^2}+\frac{\sin(\alpha-\beta)\cos\alpha}{M_H^2}+\frac{\sin\beta}{M_A^2}\Bigr),
\end{split} \end{align}
    and others are zero. 
    So, at first, we should check whether or not 
\begin{align} \begin{split}
a_+=a_-, \ \ 
b_+=b_-, \ \ 
c_\pm=d_\pm=e_\pm=0.    
\end{split} \end{align}
    If $h^0$ or $A^0$ is decoupled, we also have to check whether or not the relation 
\begin{align} \begin{split}
a_+=a_-=4b_+=4b_-     
\end{split} \end{align}
    is satisfied.
    
\begin{figure}[t]
\begin{center}
   \includegraphics[keepaspectratio=true,height=50mm]{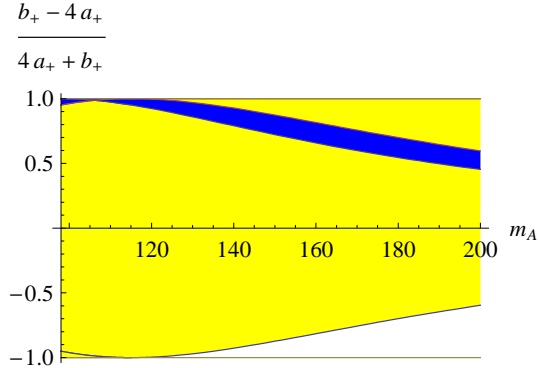}
  \caption{$M_{h^0}=98$GeV, $M_{H^0}=115$GeV, $98\le M_{A^0}\le 200$GeV. The yellow region is the allowed region in the Babu model, and the blue region is the allowed region in the Type-I\!I\!I 2HDM.}
\label{fig:2HDMMSSM-1.eps}
\end{center}
\end{figure}

\begin{figure}[t]
\begin{center}
   \includegraphics[keepaspectratio=true,height=50mm]{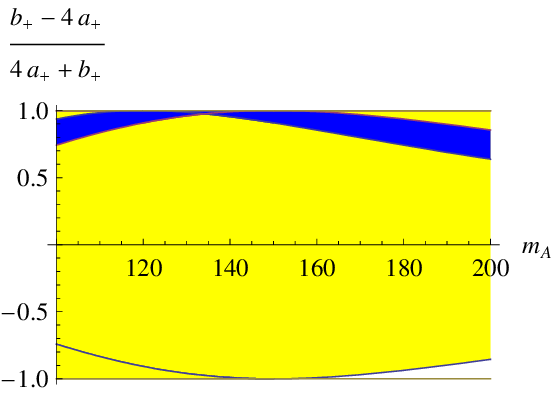}
  \caption{$M_{h^0}=120$GeV, $M_{H^0}=150$GeV, $100\le M_{A^0}\le 200$GeV. The color condition is the same as that of Fig. \ref{fig:2HDMMSSM-1.eps}.}
  \label{fig:2HDMMSSM-2.eps}
\end{center}
\end{figure}

\begin{figure}[t]
\begin{center}
   \includegraphics[keepaspectratio=true,height=50mm]{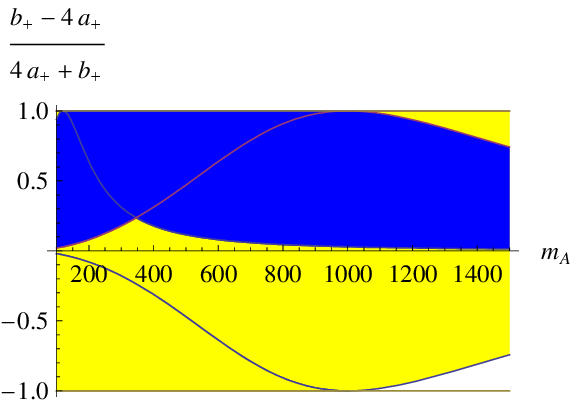}
  \caption{$M_{h^0}=120$GeV, $M_{H^0}=1000$GeV, $100\le M_{A^0}\le 1500$GeV. The color condition is the same as that of Fig. \ref{fig:2HDMMSSM-1.eps}.}
\label{fig:2HDMMSSM-3.eps}
\end{center}
\end{figure}

    Type-I\!I\!I 2HDM can also make the similar situation as BKM.
    Even if so, we have a procedure to distinguish them when both of $h^0$ and $A^0$ are not decoupled.
    In Type-I\!I\!I 2HDM, from the relation,    
\begin{align} \begin{split}
\frac{1}{M_{H^0}^2}
<\Bigl(
  \frac{\sin^2(\alpha-\beta)}{M_{H^0}^2}
 +\frac{\cos^2(\alpha-\beta)    }{M_{h^0}^2}
  \Bigr)
< \frac{1  }{M_{h^0}^2},
\end{split} \end{align}   
    the observable
\begin{align} \begin{split}    
\frac{b_+-4a_+}{b_++4a_+}
=\frac{2\Bigl(
  \frac{\sin^2(\alpha-\beta)}{M_{H^0}^2}
 +\frac{\cos^2(\alpha-\beta)    }{M_{h^0}^2}
  \Bigr)\frac{1}{M_{A^0}^2}
}{
\Bigl(
  \frac{\sin^2(\alpha-\beta)}{M_{H^0}^2}
 +\frac{\cos^2(\alpha-\beta)    }{M_{h^0}^2}
  \Bigr)^2+\Bigl(\frac{1}{M_{A^0}^2}\Bigr)^2
}    
\end{split} \end{align}    
    is constrained.    
    Similarly, in Babu model, the relation
 \begin{align} \begin{split}
-\frac{1}{M_{H^0}^2}
&\le
\Bigl(\frac{\cos(\alpha-\beta)\sin\alpha}{M_h^2}+\frac{\sin(\alpha-\beta)\cos\alpha}{M_H^2}\Bigr)\frac{1}{\sin\beta}
\\&=
\frac{(M_{H^0}^2+M_{h^0}^2) \frac{\sin(2 \alpha-\beta)}{\sin\beta}+(M_{H^0}^2-M_{h^0}^2) }{2 M_{H^0}^2 M_{h^0}^2}   
\le\infty
\end{split} \end{align} 
    constrains the observable    
\begin{align} \begin{split}
\frac{b_+-4a_+}{b_++4a_+}
=
\frac{2\Bigl(\frac{\cos(\alpha-\beta)\sin\alpha}{M_h^2}+\frac{\sin(\alpha-\beta)\cos\alpha}{M_H^2}\Bigr)\frac{1}{\sin\beta}
\frac{1}{M_A^2}
}{
\Bigl(\frac{\cos(\alpha-\beta)\sin\alpha}{M_h^2}+\frac{\sin(\alpha-\beta)\cos\alpha}{M_H^2}\Bigr)^2\Bigl(\frac{1}{\sin\beta}\Bigr)^2
+\Bigl(\frac{1}{M_A^2}\Bigr)^2}.
\end{split} \end{align}
    Figs. \ref{fig:2HDMMSSM-1.eps}, \ref{fig:2HDMMSSM-2.eps} and \ref{fig:2HDMMSSM-3.eps} are the allowed regions of this observable in BKM and Type-I\!I\!I 2HDM.
    In these figures, the yellow region is the allowed rgion in the BKM, and the blue region is the allowed rgion in the Type-I\!I\!I 2HDM.    
    Fig. \ref{fig:2HDMMSSM-1.eps} suggests the allowed regions in the condition, $M_{h^0}=98$GeV and $M_{H^0}=115$GeV.
    Similarly, Fig. \ref{fig:2HDMMSSM-2.eps} suggests the allowed regions in the condition, $M_{h^0}=120$GeV and $M_{H^0}=150$GeV; and  Fig. \ref{fig:2HDMMSSM-3.eps} suggests the allowed regions in the condition, $M_{h^0}=120$GeV and $M_{H^0}=1000$GeV.
    In each case, the allowed region in  the Type-I\!I\!I 2HDM is smaller than that of the BKM.

%%%%%%%%%%%%%%%%%%%%%%%%%%%%%%%%%%%%%%%%%%%%%%
%%%%%%%%%%%%%%%%%%%%%%%%%%%%%%%%%%%%%%%%%%%%%%
%%%%%%%%%%%%%%%%%%%%%%%%%%%%%%%%%%%%%%%%%%%%%%
%%%%%%%%%%%%%%%%%%%%%%%%%%%%%%%%%%%%%%%%%%%%%%

\section{Summary and Discussion}\label{SS5}

    We studied the energy distributions and angular distributions of $\tau\to3\mu$ decay products supposing the type-I\!I\!I 2HDM.
    We supposed four scenarios in section \ref{SS3}.
    Each scenario has the different feature which we can check using the experimental observables e.g. $a_+$, $b_+$, $a_+/b_+$, $c_+/a_+$,,,.
    We explained the upper limits on the effective coupling constants in some cases of the scenario 1.
    We suggested that the difference between the MSSM case and the type-I\!I\!I 2HDM case can be observed in energy and angular distributions.
    Also, we suggested that the difference between the BKM case and the type-I\!I\!I 2HDM case can be checked by the observable  $(b_+-4a_+)/(b_++4a_+)$.

    $\tau\to3\mu$ decay is a pure leptonic LFV event.
    Pure leptonic event has no QCD ambiguity and LFV is a null test.
    So, the experimental results become clear and sensitive to the new physics.

    The discovery of LFV event is very important, itself.
    What we want to know next is the source of it.
    This paper focus the type-I\!I\!I 2HDM as a source.
    Comparing it to the MSSM and BKM, we suggest that the energy and angular distribution are the effective probe to distinguish the models.

    Even if no new particle except for one Higgs boson is discovered in LHC, the $\tau\to3\mu$ events and its energy and angular distributions support the new physics in Higgs sector.
    High energy collider like LHC is the primary approach to the new physics study.
    Our analysis is the complimentary approach to them.

    The type-I\!I\!I 2HDM has many source of CP violation.
    This analysis manifests its effects in the observables.
    The non-zero values of $f_+$ or $g_+$ defined in Eq. (\ref{observables}) or  $(b_+\pm b_-)(c_+\mp c_-)-(e_+\mp e_-)^2$ means the CP violation.

    As a next to LHC high energy collider, the international linear collider (ILC) is planned in near future \cite{high energy process}.
    If the source of LFV is the Higgs bosons discovered in LHC, the ILC will create a number of LFV events since the Higgs bosons becomes the on-shell particle.
    However, this analysis enables us to precisely determine the next to SM before the ILC era.

%\section*{Reference}

%\chapter{Appendix}

%\chapter{Appendix}
\appendix

\section{Coupling Constants of the Model}\label{coupling constants}

$g_1$,...,$g_6$, $A_L$ and $A_R$ in Type-I\!I\!I 2HDM are written as follows:
\begin{align} \begin{split}
g_1 
&=
\frac{1}{- 2\sqrt{2}G_F}\Bigl(
 J_{23}^{H^0*} J_{22}^{H^0*} \frac{1}{M_{H^0}^2}+J_{23}^{h^0*} J_{22}^{h^0*} \frac{1}{M_{h^0}^2}-J_{23}^{A^0*} J_{22}^{A^0*} \frac{1}{M_{A^0}^2}
 \Bigr),
\\
g_2
&=
\frac{1}{- 2\sqrt{2}G_F}\Bigl(
 J_{32}^{H^0} J_{22}^{H^0}  \frac{1}{M_{H^0}^2}+J_{32}^{h^0} J_{22}^{h^0}  \frac{1}{M_{h^0}^2}-J_{32}^{A^0} J_{22}^{A^0}  \frac{1}{M_{A^0}^2}
\Bigr),
\end{split} \end{align}
\begin{align} \begin{split}
g_3 = \frac{e^2}{2\sqrt{2}G_F}
&
\biggl[\frac{(J_{13}^{H^0*}J_{12}^{H^0}+J_{23}^{H^0*}J_{22}^{H^0})(-1+3\log[\frac{q^2}{M_{H^0}^2}])+J_{33}^{H^0*}J_{32}^{H^0}(4+3\log[\frac{m_\tau^2}{M_{H^0}^2}]) }
{9(4\pi)^2 M_{H^0}^2}
\\&+(H^0\to h^0)+(H^0\to A^0)-\frac{(J^{H^+\dagger} J^{H^+})_{32}}{9(4\pi)^2 M_{H^+}^2}\biggr],
\\\\
g_4 = \frac{e^2}{2\sqrt{2}G_F}
&\biggl[\frac{(J_{31}^{H^0}J_{21}^{H^0*}+J_{32}^{H^0}J_{22}^{H^0*})(-1+3\log[\frac{q^2}{M_{H^0}^2}])+(J_{33}^{H^0}J_{23}^{H^0*})(4+3\log[\frac{m_\tau^2}{M_{H^0}^2}]) }{9(4\pi)^2 M_{H^0}^2}
\\&+(H^0\to h^0)+(H^0\to A^0)\biggr],
\end{split} \end{align}
\begin{align} \begin{split}
 g_5 
&=\frac{1}{- 4\sqrt{2}G_F}\Bigl(
 J_{23}^{H^0*} J_{22}^{H^0}  \frac{1}{M_{H^0}^2}+(H^0\to h^0)+(H^0\to A^0)\Bigr)
+g_3,
\\
g_6
 &=
 \frac{1}{- 4\sqrt{2}G_F}\Bigl(
  J_{32}^{H^0} J_{22}^{H^0*} \frac{1}{M_{H^0}^2}
+(H^0\to h^0)+(H^0\to A^0)
\Bigr)
+g_4,
\end{split} \end{align}
\begin{align} \begin{split}
A_L = \frac{e}{-4\sqrt{2}G_F m_\tau}
\Biggl[&
\frac{(J^{H^+\dagger} J^{H^+})_{32}m_\tau}{6(4\pi)^2 M_{H^+}^2}
-
\frac{(J^{H^0\dagger}J^{H^0})_{32}m_\tau}{6(4\pi)^2 M_{H^0}^2} 
-
\frac{(J^{H^0}J^{H^0\dagger})_{32}m_\mu}{6(4\pi)^2 M_{H^0}^2} 
\\&+(H^0\to h^0)+(H^0\to A^0)
\\&+
\frac{(J_{31}^{H^0}J_{12}^{H^0}m_e+J_{32}^{H^0}J_{22}^{H^0}m_\mu)}{2(4\pi)^2M_{H^0}^2}  
 (-1+2\log[\frac{q^2}{M_{H^0}^2}])
\\&+\frac{J_{33}^{H^0}J_{32}^{H^0}m_\tau}{2(4\pi)^2M_{H^0}^2}  
 (3+2\log[\frac{m_\tau^2}{M_{H^0}^2}])
+(H^0\to h^0)-(H^0\to A^0)
\Biggr],
\\\\
A_R = \frac{e}{-4\sqrt{2}G_F m_\tau}
\Biggl[&
\frac{(J^{H^+\dagger} J^{H^+})_{32}m_\mu}{6(4\pi)^2 M_{H^+}^2}
-\frac{(J^{H^0\dagger}J^{H^0})_{32} m_\mu}{6(4\pi)^2 M_{H^0}^2}
-\frac{(J^{H^0}J^{H^0\dagger})_{32}m_\tau}{6(4\pi)^2 M_{H^0}^2}
\\&+(H^0\to h^0)+(H^0\to A^0)
\\&+
\frac{(J_{13}^{H^0*}J_{21}^{H^0*}m_e+J_{23}^{H^0*}J_{22}^{H^0*}m_\mu)}{(4\pi)^2} 
 \frac{-1+2\log[\frac{q^2}{M_{H^0}^2}]}{2M_{H^0}^2}
\\&
+
\frac{J_{33}^{H^0*}J_{23}^{H^0*}m_\tau}{(4\pi)^2} 
 \frac{3+2\log[\frac{m_\tau^2}{M_{H^0}^2}]}{2M_{H^0}^2}
+(H^0\to h^0)-(H^0\to A^0)
\Biggr],
\end{split} \end{align}
    where we set $q^2=16m_\tau^2\delta^2/9\simeq2m_\mu^2$ by the condition
\begin{align} \begin{split} 
x_1<1-(\frac{4\delta}{3})^2.
\end{split} \end{align}

    Vertex Feynman rules for Type-I\!I\!I 2HDM are as follows:

\begin{tabular}{cc}
\begin{minipage}{0.355\hsize}
\begin{flushleft}
\fcolorbox{white}{white}{
  \begin{picture}(255,136) (15,-14)
    \SetWidth{0.5}
    \SetColor{Black}
    \ArrowLine(55,11)(90,46)
    \ArrowLine(90,46)(55,81)
    \DashLine(90,46)(140,46){4}
    \Text(30,96)[lb]{\Large{\Black{$\bar{\nu}_{iL}$}}}
    \Text(30,-14)[lb]{\Large{\Black{$E_{jR}$}}}
    \Text(120,56)[lb]{\Large{\Black{$H^+$}}}
    \Text(150,38)[lb]{\Large{\Black{%$=[\nu_{iL}^\dagger E_{jR} H^+]$
    }}}
  \end{picture}
}
\end{flushleft}
\end{minipage}
\begin{minipage}{0.57\hsize}
\begin{eqnarray}
 \begin{split}
&
\\&
\\&\equiv 
\sqrt{2}iJ_{ij}^{H^+}P_R
\\&=
-\frac{ig m_i \delta_{ij}}{\sqrt{2}\sin\beta M_W}\cos\beta P_R+\frac{i\eta_{ij}^E}{\sin\beta}P_R,
\end{split} \end{eqnarray}
\end{minipage}
\end{tabular}

\begin{tabular}{cc}
\begin{minipage}{0.355\hsize}
\begin{center}
\fcolorbox{white}{white}{
  \begin{picture}(255,136) (15,-14)
    \SetWidth{0.5}
    \SetColor{Black}
    \ArrowLine(55,11)(90,46)
    \ArrowLine(90,46)(55,81)
    \DashLine(90,46)(140,46){4}
    \Text(30,96)[lb]{\Large{\Black{$\bar{E}_{jR}$}}}
    \Text(30,-14)[lb]{\Large{\Black{$\nu_{iL}$}}}
    \Text(120,56)[lb]{\Large{\Black{$H^-$}}}
    \Text(150,38)[lb]{\Large{\Black{%$=[\nu_{iL} E_{jR}^\dagger H^-]$
    }}}
  \end{picture}
}
\end{center}
\end{minipage}
\begin{minipage}{0.57\hsize}
\begin{eqnarray}
\begin{split} 
&
\\&
\\&\equiv
\sqrt{2}iJ_{ij}^{H^+*}P_L
\\&=
-\frac{ig m_i \delta_{ij}}{\sqrt{2}\sin\beta M_W}\cos\beta P_L+\frac{i\eta_{ij}^E*}{\sin\beta}P_L,
\end{split} 
\end{eqnarray}
\end{minipage}
\end{tabular}

\begin{tabular}{cc}
\begin{minipage}{0.38\hsize}
\begin{center}
\fcolorbox{white}{white}{
  \begin{picture}(255,136) (15,-14)
    \SetWidth{0.5}
    \SetColor{Black}
    \ArrowLine(55,11)(90,46)
    \ArrowLine(90,46)(55,81)
    \DashLine(90,46)(140,46){4}
    \Text(30,96)[lb]{\Large{\Black{$\bar{\ell}_{iL}$}}}
    \Text(30,-14)[lb]{\Large{\Black{$\ell_{jR}$}}}
    \Text(96,76)[lb]{\Large{\Black{$\{H^0,h^0,A^0\}$}}}
    \Text(150,38)[lb]{\Large{\Black{%$=[\ell_{iL}^\dagger \ell_{jR} \{H^0,h^0,A^0\}]$
    }}}
  \end{picture}
}
\end{center}
\end{minipage}
\begin{minipage}{0.545\hsize}
\begin{align} \begin{split}\\
\\
\\
\\
\\
%[\ell_{iL}^\dagger \ell_{jR} \{H^0,h^0,A^0\}]
\equiv
iJ_{ij}^{H^0}P_R
=
-\frac{ig m_i \delta_{ij}}{2\sin\beta M_W}\sin\alpha P_R+\frac{i\eta_{ij}^E}{\sqrt{2}\sin\beta}\sin(\alpha-\beta)P_R
\\
\phantom{=},
iJ_{ij}^{h^0}P_R=
-\frac{ig m_i \delta_{ij}}{2\sin\beta M_W}\cos\alpha P_R+\frac{i\eta_{ij}^E}{\sqrt{2}\sin\beta}\cos(\alpha-\beta)P_R
\\
\phantom{=},
J_{ij}^{A^0}P_R=
\frac{g m_i \delta_{ij}}{2\sin\beta M_W}\cos\beta P_R+\frac{\eta_{ij}^E}{\sqrt{2}\sin\beta}P_R,
\phantom{=aaaaaaaaa}
\end{split} \end{align}\end{minipage}
\end{tabular}

\begin{tabular}{cc}
\begin{minipage}{0.38\hsize}
\begin{center}
\fcolorbox{white}{white}{
  \begin{picture}(255,136) (15,-14)
    \SetWidth{0.5}
    \SetColor{Black}
    \ArrowLine(55,11)(90,46)
    \ArrowLine(90,46)(55,81)
    \DashLine(90,46)(140,46){4}
    \Text(30,96)[lb]{\Large{\Black{$\bar{\ell}_{jR}$}}}
    \Text(30,-14)[lb]{\Large{\Black{$\ell_{iL}$}}}
    \Text(96,76)[lb]{\Large{\Black{$\{H^0,h^0,A^0\}$}}}
    \Text(150,38)[lb]{\Large{\Black{%$=[\ell_{jR}^\dagger \ell_{iL} \{H^0,h^0,A^0\}]$
    }}}
  \end{picture}
}
\end{center}
\end{minipage}
\begin{minipage}{0.545\hsize}
\begin{align} \begin{split}
\\
\\
\\
\\
\\%%%%%%%%%%%%%%%%%%%%%%%%%%%%%%%
%[\ell_{jR}^\dagger \ell_{iL} \{H^0,h^0,A^0\}]
\equiv
iJ_{ij}^{H^0*} P_L
=
-\frac{ig m_i \delta_{ij}}{2\sin\beta M_W}\sin\alpha P_L+\frac{i\eta_{ij}^E*}{\sqrt{2}\sin\beta}\sin(\alpha-\beta)P_L
\\
\phantom{=},
iJ_{ij}^{h^0*}P_L=
-\frac{ig m_i \delta_{ij}}{2\sin\beta M_W}\cos\alpha P_L+\frac{i\eta_{ij}^E*}{\sqrt{2}\sin\beta}\cos(\alpha-\beta)P_L
\\
\phantom{=},
-J_{ij}^{A^0*}P_L=
-\frac{g m_i \delta_{ij}}{2\sin\beta M_W}\cos\beta P_L
-\frac{\eta_{ij}^E*}{\sqrt{2}\sin\beta}P_L.
\phantom{aalaaaaa}
\end{split} \end{align}
\end{minipage}
\end{tabular}

\section{Higgs masses}\label{Appendix Higgs}

    In Type-I\!I\!I 2HDM, the general Higgs potential is written as \cite{Hhg}:
\begin{align} \begin{split}
V(\phi_1,\phi_2)&=
\lambda_1(\phi^\dagger_1 \phi_1-v_1^2)^2+\lambda_2(\phi^\dagger_2 \phi_2-v_2^2)^2
\\&+\lambda_3[(\phi^\dagger_1 \phi_1-v_1^2)+(\phi^\dagger_2 \phi_2-v_2^2)]^2
\\&+\lambda_4[(\phi^\dagger_1 \phi_1)(\phi^\dagger_2 \phi_2)-(\phi^\dagger_1 \phi_2)(\phi^\dagger_2 \phi_1)]
\\&+\lambda_5[Re(\phi^\dagger_1 \phi_2)-v_1 v_2 \cos \xi]^2
\\&+\lambda_6[Im(\phi^\dagger_1 \phi_2)-v_1 v_2 \sin \xi]^2,
\end{split} \end{align},
    where $\lambda_1,...,\lambda_6$ are the real coefficients;
    $\phi_1$ and $\phi_2$ are the Higgs fields with the vacume expectation values, $v_1$ and $v_2$, respectively; and $\xi$ is the CP phase.
%    and $\tan\beta=v_2/v_1$.
%
%
    When we set $\xi=0$ for CP conservation of Higgs potential, the neutral CP even and odd Higgs masses are written as
\begin{align} \begin{split}
M_{H^0,h^0}&=\frac{1}{2}[M_{11}+M_{22}\pm\sqrt{(M_{11}-M_{22})^2+4M_{12}^2}],\\
 M_{A^0}^2&=\lambda_6(v_1^2+v_2^2),
\end{split} \end{align}
    respectively, where 
\begin{align} \begin{split}
  M_{11}&=4v_1^2(\lambda_1+\lambda_3)+v_2^2\lambda_5,
\\M_{22}&=4v_2^2(\lambda_2+\lambda_3)+v_1^2\lambda_5,
\\M_{12}&=M_{21}=(4\lambda_3+\lambda_5)v_1 v_2.
\end{split} \end{align}
    We define the mixing angle $\alpha$ as
\begin{align} \begin{split}
\tan 2 \alpha =\frac{2M_{12}}{M_{11}-M_{22}}.
\end{split} \end{align}

\section{Radiative one-loop diagrams}\label{Appendix calc}

    We explain the general charged fermion-charged fermion-photon one-loop diagrams which contain scalar fields in loop.

    For convenience, we define the projection operators,
\begin{align} \begin{split}
P_1=\frac{1\pm\gamma_5}{2},\ \ \ \ 
P_2=\frac{1\mp\gamma_5}{2}.
\end{split} \end{align}

\subsection{Neutral Scalar}

%$$$$$$$$$$$$$$$$$$$$$$$$$$$$$$$$$$$$$$$
%$$$$$$$$$$$$$$$$$$$$$$$$$$$$$$$$$$$$$$$
%$$$$$$$$$$$$$$$$$$$$$$$$$$$$$$$$$$$$$$$
%$$$$$$$$$$$$$$$$$$$$$$$$$$$$$$$$$$$$$$$
%$$$$$$$$$$$$$$$$$$$$$$$$$$$$$$$$$$$$$$$
%$$$$$$$$$$$$$$$$$$$$$$$$$$$$$$$$$$$$$$$

    We set the general charged fermion-charged fermion-neutral scalor and charged fermion-charged fermion-photon vertices as follows:
\begin{center}
\fcolorbox{white}{white}{
  \begin{picture}(255,136) (15,-14)
    \SetWidth{0.5}
    \SetColor{Black}
    \ArrowLine(55,11)(90,46)
    \ArrowLine(90,46)(55,81)
    \DashLine(90,46)(140,46){4}
    \Text(30,96)[lb]{\Large{\Black{$\bar{\ell}_{i2}$}}}
    \Text(30,-14)[lb]{\Large{\Black{${\ell_k}_1$}}}
    \Text(120,56)[lb]{\Large{\Black{$h^0$}}}
    \Text(150,38)[lb]{\Large{\Black{$=iV_{ik}^1 P_1,$}}}
  \end{picture}
}
\end{center}

\begin{center}
\fcolorbox{white}{white}{
  \begin{picture}(255,136) (15,-14)
    \SetWidth{0.5}
    \SetColor{Black}
    \ArrowLine(55,11)(90,46)
    \ArrowLine(90,46)(55,81)
    \DashLine(90,46)(140,46){4}
    \Text(30,96)[lb]{\Large{\Black{$\bar{\ell}_{i1}$}}}
    \Text(30,-14)[lb]{\Large{\Black{${\ell_k}_2$}}}
    \Text(120,56)[lb]{\Large{\Black{$h^0$}}}
    \Text(150,38)[lb]{\Large{\Black{$=iV_{ik}^2 P_2,$}}}
  \end{picture}
}
\end{center}

\begin{center}
\fcolorbox{white}{white}{
  \begin{picture}(255,136) (15,-14)
    \SetWidth{0.5}
    \SetColor{Black}
    \ArrowLine(55,11)(90,46)
    \ArrowLine(90,46)(55,81)
    \Photon(90,46)(140,46){6.5}{3}
    \Text(30,96)[lb]{\Large{\Black{${\bar{\ell}_i}$}}}
    \Text(30,-14)[lb]{\Large{\Black{${\ell_k}$}}}
    \Text(120,56)[lb]{\Large{\Black{$\gamma^\mu$}}}
    \Text(150,38)[lb]{\Large{\Black{$=ieQ \gamma^\mu \delta_{ik}$,}}}
  \end{picture}
}
\end{center}
    where $\ell$ is the charged fermion; 
    $\ell_{ir}=P_r\ell_{i}$, where $r=1,2$; 
    $h^0$ is the neutral scalar; 
    $V_{ik}^r$ are the complex coupling constant;
     and $\gamma^\mu$ is the photon; $Q=-1$ for a lepton.
     
%%%%%%%%%%%%%%%%%%%%%%%%%%%%%%%%%%%%%%%%%%%%%%%%%%%%%%%%%%%%%%%%%
%%%%%%%%%%%%%%%%%%%%%%%%%%%%%%%%%%%%%%%%%%%%%%%%%%%%%%%%%%%%%%%%%

    The one-loop diagrams which contain the neutral scalars are written as:
\begin{center}
\fcolorbox{white}{white}{
  \begin{picture}(240,211) (30,-14)
    \SetWidth{0.5}
    \SetColor{Black}
    \ArrowArcn(150,121)(30,0,-90)
    \ArrowArcn(150,121)(30,-90,-180)
    \DashCArc(150,121)(30,-0,180){2}
    \ArrowLine(120,121)(45,121)
    \ArrowLine(255,121)(180,121)
    \Text(30,146)[lb]{\Large{\Black{$\bar{\ell}_i^+(p)$}}}
    \Text(210,146)[lb]{\Large{\Black{$\ell_j^+(p-q)$}}}
    \Text(135,176)[lb]{\Large{\Black{$h^0(p-k)$}}}
    \Text(185,76)[lb]{\Large{\Black{$\ell_k(k-q)$}}}
    \Photon(150,91)(150,16){7.5}{4}
    \Text(150,-14)[lb]{\Large{\Black{$\gamma^*_\mu(q)$}}}
    \LongArrow(135,61)(135,31)
    \LongArrow(105,106)(120,91)
    \LongArrow(180,91)(195,106)
    \Text(90,76)[lb]{\Large{\Black{$\ell_k(k)$}}}
    \Text(305,121)[lb]{\Large{\Black{$=i\mathcal{M}_1,$}}}
  \end{picture}
}
\end{center}

\begin{center}
\fcolorbox{white}{white}{
  \begin{picture}(240,181) (30,-14)
    \SetWidth{0.5}
    \SetColor{Black}
    \ArrowArcn(150,91)(30,0,-180)
    \DashCArc(150,91)(30,-0,180){2}
    \ArrowLine(120,91)(45,91)
    \ArrowLine(255,91)(180,91)
    \Text(30,116)[lb]{\Large{\Black{$\bar{\ell}_i^+(p)$}}}
    \Text(210,116)[lb]{\Large{\Black{$\ell_j^+(p-q)$}}}
    \Text(135,131)[lb]{\Large{\Black{$h^0(p-k)$}}}
    \Photon(75,91)(75,16){7.5}{4}
    \Text(75,-14)[lb]{\Large{\Black{$\gamma^*_\mu(q)$}}}
    \LongArrow(60,61)(60,31)
    \LongArrow(135,46)(165,46)
    \Text(150,16)[lb]{\Large{\Black{$\ell_k(k-q)$}}}
    \LongArrow(90,106)(105,106)
    \Text(75,121)[lb]{\Large{\Black{$\ell_i(p-q)$}}}
    \Text(305,91)[lb]{\Large{\Black{$=i\mathcal{M}_2,$}}}
  \end{picture}
}
\end{center}

\begin{center}
\fcolorbox{white}{white}{
  \begin{picture}(240,181) (30,-14)
    \SetWidth{0.5}
    \SetColor{Black}
    \ArrowArcn(150,91)(30,0,-180)
    \DashCArc(150,91)(30,-0,180){2}
    \ArrowLine(120,91)(45,91)
    \ArrowLine(255,91)(180,91)
    \Text(30,121)[lb]{\Large{\Black{$\bar{\ell}_i^+(p)$}}}
    \Text(240,121)[lb]{\Large{\Black{$\ell_j^+(p-q)$}}}
    \Text(135,131)[lb]{\Large{\Black{$h^0(p-k)$}}}
    \Photon(225,91)(225,16){7.5}{4}
    \Text(225,-14)[lb]{\Large{\Black{$\gamma^*_\mu(q)$}}}
    \Text(150,21)[lb]{\Large{\Black{$\ell_k(k)$}}}
    \LongArrow(240,61)(240,31)
    \LongArrow(135,46)(165,46)
    \LongArrow(195,106)(210,106)
    \Text(240,121)[lb]{\Large{\Black{$\ell_j^+(p-q)$}}}
    \Text(195,111)[lb]{\Large{\Black{$\ell_j(p)$}}}
    \Text(305,91)[lb]{\Large{\Black{$=i\mathcal{M}_3$.}}}
  \end{picture}
}
\end{center}

%$$$$$$$$$$$$$$$$$$$$$$$$$$$$$$$$$$$$$$$
%$$$$$$$$$$$$$$$$$$$$$$$$$$$$$$$$$$$$$$$
%$$$$$$$$$$$$$$$$$$$$$$$$$$$$$$$$$$$$$$$
%$$$$$$$$$$$$$$$$$$$$$$$$$$$$$$$$$$$$$$$
%$$$$$$$$$$$$$$$$$$$$$$$$$$$$$$$$$$$$$$$
%$$$$$$$$$$$$$$$$$$$$$$$$$$$$$$$$$$$$$$$

\subsubsection{Neutral Higgs Contribution with $iV_{ik}^1 P_1$ and $iV_{kj}^1 P_1$}

\begin{align} \begin{split}\label{n11}
i\mathcal{M}_1^\mu+i\mathcal{M}_2^\mu&+i\mathcal{M}_3^\mu\Bigr|_{11}
\\=&
\frac{-ieQV_{ik}^1V_{kj}^1 m_k}{(4\pi)^2}\int dx \int dy \bar{v}(p)
\frac{1}{\Delta}\biggl[
 (y-1)i\sigma^{\nu\mu}q_\nu P_1 \\
&+\frac{(2x+y-1)q^2}{m_i^2-m_j^2}\gamma^\mu(m_i P_1+m_j P_2)
+(2x+y-1)q^\mu P_1\biggr]
v(p-q),
\end{split} \end{align}
    where
\begin{align} \begin{split}
\Delta&  =x(x+y-1)q^2+y(x+y-1)m_i^2-xym_j^2+(1-y)m_k^2+yM^2;
\end{split} \end{align}
    $|_{11}$ in the left hand side of Eq. (\ref{n11}) means that we use $iV_{ik}^1 P_1$ and $iV_{kj}^1 P_1$ vertices;
    $m_i$, $m_j$ and $m_k$ are the $\ell_i$, $\ell_j$ and $\ell_k$ masses, respectively;
    and $M$ is the $h^0$ mass.

\subsubsection{Neutral Higgs Contribution with $iV_{ik}^2 P_2$ and $iV_{kj}^1 P_1$}

\begin{align} \begin{split}
i\mathcal{M}_1^\mu+i\mathcal{M}_2^\mu&+i\mathcal{M}_3^\mu\Bigr|_{21}
\\=&
\frac{-ieQV_{ik}^2V_{kj}^1}{(4\pi)^2}\int dx dy \frac{1}{\Delta}\bar{v}(p)
\biggl[
\\&q^2\gamma^\mu\frac{1}{m_j^2-m_i^2}
\Bigl\{x(2x+y-2)m_j^2 P_1-(x+y-1)(2x+y)m_i^2 P_1-y(y+2x-1)m_i m_j P_2\Bigr\}
\\&+(x+y-1)(2x+y)m_i q^\mu P_1-x(2x+y-2)m_j q^\mu P_2
\\&-i\sigma^{\nu\mu}q_\nu \bigl\{xym_j P_2-y(x+y-1)m_i P_1\bigr\}
\biggr]
v(p-q),
\end{split} \end{align}
    where $|_{21}$ in the left hand side means that we use $iV_{ik}^2 P_2$ and $iV_{kj}^1 P_1$ vertices.

\subsection{charged scalar}

%$$$$$$$$$$$$$$$$$$$$$$$$$$$$$$$$$$$$$$$
%$$$$$$$$$$$$$$$$$$$$$$$$$$$$$$$$$$$$$$$
%$$$$$$$$$$$$$$$$$$$$$$$$$$$$$$$$$$$$$$$
%$$$$$$$$$$$$$$$$$$$$$$$$$$$$$$$$$$$$$$$
%$$$$$$$$$$$$$$$$$$$$$$$$$$$$$$$$$$$$$$$
%$$$$$$$$$$$$$$$$$$$$$$$$$$$$$$$$$$$$$$$

    We set the general charged fermion-neutral fermion-charged scalor and charged scalar-charged scalar-photon vertices as follows:
\begin{center}
\fcolorbox{white}{white}{
  \begin{picture}(255,136) (15,-14)
    \SetWidth{0.5}
    \SetColor{Black}
    \ArrowLine(55,11)(90,46)
    \ArrowLine(90,46)(55,81)
    \DashLine(90,46)(140,46){4}
    \Text(30,96)[lb]{\Large{\Black{$\bar{\ell}_{i2}$}}}
    \Text(30,-14)[lb]{\Large{\Black{${\nu_k}_1$}}}
    \Text(120,56)[lb]{\Large{\Black{$H^-$}}}
    \Text(150,38)[lb]{\Large{\Black{$=iW_{ik}^1 P_1,$}}}
  \end{picture}
}
\end{center}

\begin{center}
\fcolorbox{white}{white}{
  \begin{picture}(255,136) (15,-14)
    \SetWidth{0.5}
    \SetColor{Black}
    \ArrowLine(55,11)(90,46)
    \ArrowLine(90,46)(55,81)
    \DashLine(90,46)(140,46){4}
    \Text(30,96)[lb]{\Large{\Black{$\bar{\ell}_{i1}$}}}
    \Text(30,-14)[lb]{\Large{\Black{${\nu_k}_2$}}}
    \Text(120,56)[lb]{\Large{\Black{$H^-$}}}
    \Text(150,38)[lb]{\Large{\Black{$=iW_{ik}^2 P_2,$}}}
  \end{picture}
}
\end{center}

\begin{center}
\fcolorbox{white}{white}{
  \begin{picture}(255,136) (15,-14)
    \SetWidth{0.5}
    \SetColor{Black}
    \ArrowLine(55,11)(90,46)
    \ArrowLine(90,46)(55,81)
    \DashLine(90,46)(140,46){4}
    \Text(30,96)[lb]{\Large{\Black{$\bar{\nu}_{i2}$}}}
    \Text(30,-14)[lb]{\Large{\Black{${\ell_k}_1$}}}
    \Text(120,56)[lb]{\Large{\Black{$H^+$}}}
    \Text(150,38)[lb]{\Large{\Black{$=iU_{ik}^1 P_1,$}}}
  \end{picture}
}
\end{center}

\begin{center}
\fcolorbox{white}{white}{
  \begin{picture}(255,136) (15,-14)
    \SetWidth{0.5}
    \SetColor{Black}
    \ArrowLine(55,11)(90,46)
    \ArrowLine(90,46)(55,81)
    \DashLine(90,46)(140,46){4}
    \Text(30,96)[lb]{\Large{\Black{$\bar{\nu}_{i1}$}}}
    \Text(30,-14)[lb]{\Large{\Black{${\ell_k}_2$}}}
    \Text(120,56)[lb]{\Large{\Black{$H^+$}}}
    \Text(150,38)[lb]{\Large{\Black{$=iU_{ik}^2 P_2,$}}}
  \end{picture}
}
\end{center}

\begin{center}
\fcolorbox{white}{white}{
  \begin{picture}(255,136) (15,-14)
    \SetWidth{0.5}
    \SetColor{Black}
    \DashLine(55,11)(90,46){4}
    \DashLine(55,81)(90,46){4}
    \LongArrow(65,11)(85,31)
    \LongArrow(65,81)(85,61)
    \Photon(90,46)(140,46){6.5}{3}
    \Text(80,6)[lb]{\Large{\Black{$p_+$}}}
    \Text(80,76)[lb]{\Large{\Black{$p_-$}}}
    \Text(30,96)[lb]{\Large{\Black{$H^-$}}}
    \Text(30,-14)[lb]{\Large{\Black{$H^+$}}}
    \Text(120,56)[lb]{\Large{\Black{$\gamma^\mu$}}}
    \Text(150,38)[lb]{\Large{\Black{$=ieQ (p_+-p_-)^\mu$,}}}
  \end{picture}
}
\end{center}
    where $H^\pm$ are the charged scalars which have $\pm1$ electromagnetic charges, respectively;
    $\nu_k$ are the neutral fermions. 
     and $p_\pm$ are the $H^\pm$ momenta, respectively.
     
%%%%%%%%%%%%%%%%%%%%%%%%%%%%%%%%%%%%%%%%%%%%%%%%%%%%%%%%%%%%%%%%%
%%%%%%%%%%%%%%%%%%%%%%%%%%%%%%%%%%%%%%%%%%%%%%%%%%%%%%%%%%%%%%%%%

    The one-loop diagrams which contain the charged scalars are
\begin{center}
\fcolorbox{white}{white}{
  \begin{picture}(240,211) (30,-14)
    \SetWidth{0.5}
    \SetColor{Black}
    \ArrowArc(150,121)(30,0,-180)
    \DashCArc(150,121)(30,180,0){2}
    \ArrowLine(120,121)(45,121)
    \ArrowLine(255,121)(180,121)
    \Text(30,146)[lb]{\Large{\Black{$\bar{\ell}_i^+(p)$}}}
    \Text(210,146)[lb]{\Large{\Black{$\ell_j^+(p-q)$}}}
    \Text(135,176)[lb]{\Large{\Black{$\nu_k(p-k)$}}}
    \Text(185,76)[lb]{\Large{\Black{$H(k-q)$}}}
    \Photon(150,91)(150,16){7.5}{4}
    \Text(150,-14)[lb]{\Large{\Black{$\gamma^*_\mu(q)$}}}
    \LongArrow(135,61)(135,31)
    \LongArrow(105,106)(120,91)
    \LongArrow(180,91)(195,106)
    \Text(90,76)[lb]{\Large{\Black{$H(k)$}}}
    \Text(305,121)[lb]{\Large{\Black{$=i\mathcal{M}_1',$}}}
  \end{picture}
}
\end{center}

\begin{center}
\fcolorbox{white}{white}{
  \begin{picture}(240,181) (30,-14)
    \SetWidth{0.5}
    \SetColor{Black}
    \ArrowArc(150,91)(30,0,-180)
    \DashCArc(150,91)(30,180,0){2}
    \ArrowLine(120,91)(45,91)
    \ArrowLine(255,91)(180,91)
    \Text(30,116)[lb]{\Large{\Black{$\bar{\ell}_i^+(p)$}}}
    \Text(210,116)[lb]{\Large{\Black{$\ell_j^+(p-q)$}}}
    \Text(135,131)[lb]{\Large{\Black{$\nu_k(p-k)$}}}
    \Photon(75,91)(75,16){7.5}{4}
    \Text(75,-14)[lb]{\Large{\Black{$\gamma^*_\mu(q)$}}}
    \LongArrow(60,61)(60,31)
    \LongArrow(135,46)(165,46)
    \Text(150,16)[lb]{\Large{\Black{$H(k-q)$}}}
    \LongArrow(90,106)(105,106)
    \Text(75,121)[lb]{\Large{\Black{$\ell_i(p-q)$}}}
    \Text(305,91)[lb]{\Large{\Black{$=i\mathcal{M}_2',$}}}
  \end{picture}
}
\end{center}

\begin{center}
\fcolorbox{white}{white}{
  \begin{picture}(240,181) (30,-14)
    \SetWidth{0.5}
    \SetColor{Black}
    \ArrowArc(150,91)(30,0,-180)
    \DashCArc(150,91)(30,180,0){2}
    \ArrowLine(120,91)(45,91)
    \ArrowLine(255,91)(180,91)
    \Text(30,121)[lb]{\Large{\Black{$\bar{\ell}_i^+(p)$}}}
    \Text(135,131)[lb]{\Large{\Black{$\nu_k(p-k)$}}}
    \Photon(225,91)(225,16){7.5}{4}
    \Text(225,-14)[lb]{\Large{\Black{$\gamma^*_\mu(q)$}}}
    \Text(150,21)[lb]{\Large{\Black{$H(k)$}}}
    \LongArrow(240,61)(240,31)
    \LongArrow(135,46)(165,46)
    \Text(240,121)[lb]{\Large{\Black{$\ell_j^+(p-q)$}}}
    \Text(195,106)[lb]{\Large{\Black{$\ell_j(p)$}}}
    \Text(305,91)[lb]{\Large{\Black{$=i\mathcal{M}_3'$.}}}
  \end{picture}
}
\end{center}

%$$$$$$$$$$$$$$$$$$$$$$$$$$$$$$$$$$$$$$$
%$$$$$$$$$$$$$$$$$$$$$$$$$$$$$$$$$$$$$$$
%$$$$$$$$$$$$$$$$$$$$$$$$$$$$$$$$$$$$$$$
%$$$$$$$$$$$$$$$$$$$$$$$$$$$$$$$$$$$$$$$
%$$$$$$$$$$$$$$$$$$$$$$$$$$$$$$$$$$$$$$$
%$$$$$$$$$$$$$$$$$$$$$$$$$$$$$$$$$$$$$$$

\subsubsection{Charged Higgs Contribution with $iW_{ik}^1 P_1$ and $iU_{kj}^1 P_1$}

\begin{align} \begin{split}\label{c11}
i\mathcal{M}_1^{'\mu}+i\mathcal{M}_2^{'\mu}&+i\mathcal{M}_3^{'\mu}\Bigr|_{11}
\\=&
\frac{-ieQW_{ik}^1U_{kj}^1 m_k}{(4\pi)^2}\int dx \int dy \bar{v}(p)
\frac{1}{\Delta'}\biggl[
 yi\sigma^{\nu\mu}q_\nu P_1 \\
&+\frac{(2x+y-1)q^2}{m_i^2-m_j^2}\gamma^\mu(m_i P_1+m_j P_2)
+(2x+y-1)q^\mu P_1\biggr]
v(p-q),
\end{split} \end{align}
    where
\begin{align} \begin{split}
\Delta'&  =x(x+y-1)q^2+y(x+y-1)m_i^2-xym_j^2+(1-y)M^2+ym_k^2;
\end{split} \end{align}
    $M$ is the $H^\pm$ mass;
    and $|_{11}$ in the left hand side of Eq. (\ref{c11}) means that we use $iW_{ik}^1 P_1$ and $iU_{kj}^1 P_1$ vertices.

\subsubsection{Charged Higgs Contribution with $iW_{ik}^2 P_2$ and $iU_{kj}^1 P_1$}

\begin{align} \begin{split}
i\mathcal{M}_1^{'\mu}+i\mathcal{M}_2^{'\mu}&+i\mathcal{M}_3^{'\mu}\Bigr|_{21}
\\=&
\frac{-ieQW_{ik}^2U_{kj}^1}{(4\pi)^2}\int dx dy \frac{1}{\Delta'}\bar{v}(p)
\biggl[
\\&-q^2\gamma^\mu\frac{2x+y-1}{m_j^2-m_i^2}
\Bigl\{x(m_j^2 P_1+m_i m_j P_2)-(x+y-1)(m_i^2 P_1+m_i m_j P_2)\Bigr\}
\\&-(x+y-1)(2x+y-1)m_i q^\mu P_1+x(2x+y-1)m_j q^\mu P_2
\\&+i\sigma^{\nu\mu}q_\nu \bigl\{xym_j P_2-y(x+y-1)m_i P_1\bigr\}
\biggr]
v(p-q),
\end{split} \end{align}
    where $|_{21}$ in the left hand side means that we use $iW_{ik}^2 P_2$ and $iU_{kj}^1 P_1$ vertices.

\subsection{Application to the Type-I\!I\!I 2HDM}

    For the type-I\!I\!I 2HDM radiative one-loop diagrams in which neutral Higgses propagate, we sum over four vertex substitutions written as: 
\begin{align} \begin{split}
  V_{ik}^1P_1 \to J_{3k}^{\{H^0,h^0,A^0\}}P_R
,\ \ \ \ \ \ V_{kj}^1P_1 \to J_{k2}^{\{H^0,h^0,A^0\}}P_R,
\end{split} \end{align}
\begin{align} \begin{split}
  V_{ik}^1P_1 \to J_{k3}^{\{H^0,h^0,A^0\}*}P_L
,\ \ \ \ \ \ V_{kj}^1P_1 \to J_{2k}^{\{H^0,h^0,A^0\}*}P_L,
\end{split} \end{align}
\begin{align} \begin{split}
  V_{ik}^2P_2 \to J_{k3}^{\{H^0,h^0,A^0\}*}P_L
,\ \ \ \ \ \ V_{kj}^1P_1 \to J_{k2}^{\{H^0,h^0,A^0\}}P_R,
\end{split} \end{align}
\begin{align} \begin{split}
  V_{ik}^2P_2 \to J_{3k}^{\{H^0,h^0,A^0\}}P_R
, \ \ \ \ \ \ V_{kj}^1P_1 \to J_{2k}^{\{H^0,h^0,A^0\}*}P_L.
\end{split} \end{align}
    Also, we set the masses as
\begin{align} \begin{split}
  m_i      &\to m_\tau,
\\m_j      &\to m_\tau, m_\mu, m_e,
\\m_k      &\to m_\mu,
\\M        &\to M_{H^0},M_{h^0},M_{A^0},
\end{split} \end{align}
    and take summation over the subscript $j$ and propagating Higgses $H^0,h^0,A^0$.

    Similarly, for the diagrams in which charged Higgses propagate, we substitute for the vertices as 
\begin{align} \begin{split}
  W_{ik}^2P_2 &\to \sqrt{2}J_{k3}^{H^+*}P_L,
\\U_{kj}^1P_1 &\to \sqrt{2}J_{k2}^{H^+}P_R.
\end{split} \end{align}
    Also, we set the masses as
\begin{align} \begin{split}
  m_i      &\to m_\tau,
\\m_j      &\to m_\tau, m_\mu, m_e,
\\m_k      &\to m_\mu,
\\M        &\to M_{H^\pm},
\end{split} \end{align}
    and take summation over the subscript $j$.

\end{document}